\documentclass[a4paper, 11pt, authoryear, round, numafflabel]{elsarticle}
\usepackage[normalem]{ulem}
\usepackage{graphicx}
\usepackage[usenames]{color}
\usepackage[flushleft]{threeparttable}
\DeclareGraphicsExtensions{.pdf,.png,.jpg,.mps,.eps,.ps}
\usepackage{amsmath,amssymb,gensymb}
\usepackage{color}
\usepackage{xcolor}
\usepackage{lscape}
\usepackage{psfrag}
\usepackage[colorlinks,citecolor=blue,urlcolor=blue,bookmarks=false,hypertexnames=true]{hyperref} 
\renewcommand{\arraystretch}{2}

\newcommand\kms{\ifmmode {\rm~km\ s}^{-1} \else ~km s$^{-1}$\fi}
\newcommand\Hunit{\ifmmode {\rm~km\ s}^{-1}\ {\rm Mpc}^{-1}
	\else ~km s$^{-1}$ Mpc$^{-1}$\fi}
\newcommand\ctssec{\ifmmode {\rm~count\ s}^{-1} \else ~count s$^{-1}$\fi}
\newcommand\ergsec{\ifmmode {\rm~erg\ s}^{-1} \else
	~erg s$^{-1}$\fi}
\newcommand\funit{\ifmmode {\rm~erg\ s}^{-1}\;{\rm cm}^{-2} \else
	~ergs s$^{-1}$ cm$^{-2}$\fi}
\newcommand\phflux{\ifmmode {\rm~photon\ s}^{-1}\;{\rm cm}^{-2}
	\else   ~photon s$^{-1}$ cm$^{-2}$\fi}
\newcommand\efluxA{\ifmmode {\rm~erg\ s}^{-1}\;{\rm cm}^{-2}\;{\rm
		\AA}^{-1} \else ~erg s$^{-1}$ cm$^{-2}$ \AA$^{-1}$\fi}
\newcommand\efluxHz{\ifmmode {\rm~erg\ s}^{-1}\;{\rm cm}^{-2}\;{\rm
		Hz}^{-1} \else ~erg s$^{-1}$ cm$^{-2}$ Hz$^{-1}$\fi}
\newcommand\cc{\ifmmode {\rm~cm}^{-3} \else cm$^{-3}$\fi}
\newcommand\FWHM{\ifmmode {\rm~FWHM} \else ${\rm~FWHM}$\fi}
\newcommand\Msun{\ifmmode M_{\odot} \else $M_{\odot}$\fi}
\newcommand\Lsun{\ifmmode L_{\odot} \else $L_{\odot}$\fi}

\newcommand\hbeta{\ifmmode {\rm H}\beta \else H$\beta$\fi}
\newcommand\Kalpha{\ifmmode {\rm K}\alpha \else K$\alpha$\fi}
\newcommand\nh{\ifmmode N_{\rm H} \else N$_{\rm H}$\fi}

\begin{document}
	
	\begin{frontmatter}
		\title{On the assessment of the disk truncation and detection of type-II-like bursts from the accreting millisecond X-ray Pulsar IGR J17062--6143}
		
		\author[1]{Aditya S. Mondal\corref{cor1}}
		\cortext[cor1]{E-mail: adityas.mondal@visva-bharati.ac.in}
		
		\author[1]{Mahasweta Bhattacharya} 
		\author[1]{Swarnendu Jana}
		\author[1]{Biplab Raychaudhuri} 
		\author[2]{Gulab C. Dewangan} 
		
		\address[1]{Department of Physics, Visva-Bharati, Santiniketan, West Bengal, 731235, India}
		\address[2]{Inter-University Centre for  Astronomy \& Astrophysics (IUCAA), Pune, 411007, India}
		
		\begin{abstract}
			
			We present a spectral analysis of the \textit{NuSTAR} and \textit{NICER} observations of the accreting millisecond X-ray pulsar IGR J17062-6143, performed in 2022. The source remained in the hard spectral state during these observations, with a luminosity of about $\sim$0.13-0.14 percent of the Eddington luminosity. The continuum emission of the \textit{NuSTAR} spectrum is entirely dominated by a power-law component or by Comptonized emission of disc photons by a plasma with a high electron temperature ($\gtrsim100$ keV). The \textit{NuSTAR} spectrum also reveals clear evidence of disc reflection, a broad Fe K line around 6-8 keV, and a Compton hump peaking at 20 keV, irrespective of the choice of the continuum models. Our spectral studies suggest a disk extending close to the neutron star surface ($\sim$6-20 $R_{\rm g}$) at low inclination angle ($\sim41$ degrees), as revealed by the self-consistent relativistic reflection model, \texttt{relxill}. In addition, we detected type-II-like bursts for the first time in the \textit{NICER} observation of this source. Light curve profiles of type-II-like bursts exhibit different patterns, mostly associated with the so-called mode-0 and transition from mode-0 to mode-1 type-II bursts. The energy spectra of the persistent (pre-burst) and burst emission are well described by an absorbed Comptonization component, scattering disk- and blackbody-distributed photons, respectively, by a corona with a temperature of 1-3 keV. Although the origin of the type-II-like burst is not very clear, it has been possibly linked to disk-magnetosphere interaction of the accretion flow. 
			
		\end{abstract}
		
	\end{frontmatter}
	
	\section{Introduction} 
	
	Neutron Star Low mass X-ray binaries (NS LMXBs) consist of a neutron star (NS) that accretes gas from a companion having less mass than the compact primary (\citealt{shapiro1983black, 1997LNP...487....1T, Bahramian_2022}). The accreted matter transferred from the donor generally possesses angular momentum and does not fall directly on the compact object \citep{lynden1974evolution}. In a binary system, mass transfer occurs when the donor star fills its Roche lobe, and matter flows through the inner Lagrangian point ($L_1$) \citep{roche1849memoire, kopal1989roche}. Due to conservation of angular momentum at $L_1$, the accreted matter forms an accretion disk around the compact object \citep{1980A&A....83..133A, frank2002accretion}.  The infalling matter in the accretion disk exerts a ram pressure as a consequence of its inward motion through the disk plasma \citep{gunn1972infall}. Neutron stars typically possess magnetic field $B\sim10^8-10^9$ G \citep{1993ApJ...408..194T}. The accretion flow towards the neutron star gets truncated at the Alfv$\acute{e}$n radius ($R_{\rm A}$), where the inward ram pressure of the accreting matter is comparable to the magnetic pressure \citep{ghosh1977accretion}. Inside the Alfv$\acute{e}$n radius, the magnetic field of the neutron star determines the flow of matter. The magnetic field channels the matter along the magnetic field lines towards the magnetic poles of the neutron star.
	The accreted matter, while falling inward to the magnetic poles of the NS, gets heated up in the due process of acceleration and radiates in X-rays. Accreting millisecond X-ray pulsars (AMXPs) are NS LMXBs where the NS spins up rapidly to high frequency ($\nu\ge100$ Hz) and the accreted gas gets channeled out of the accretion disk by weak magnetic fields (B$\sim$10$^{8-9}$ G) onto the magnetic poles of the neutron star giving rise to the X-ray pulsations at the spin frequency \citep{2021ASSL..461..143P}.\\
	
	The extent to which the inner accretion disk can penetrate toward the neutron star is governed by the magnetic field of the neutron star \citep{lamb1973model}. The accretion disk is truncated near magnetospheric radius ($R_{\rm m}$), which is a fraction of the Alfv$\acute{e}$n radius, $R_{\rm m}= \xi R_{\rm A}$, where $\xi(\sim0.5-1)$ depends on the geometry and disk-magnetosphere interaction \citep{ghosh1979accretiona}. At the magnetospheric radius, the ram pressure of the accreting matter becomes comparable to the magnetic pressure of the NS magnetosphere. The co-rotation radius ($R_{\rm co}$) is defined as the distance from the neutron star where the angular velocity of matter in a Keplerian disk is equal to the spin angular frequency of the neutron star. Inside the magnetosphere, the accretion flow departs from Keplerian motion and becomes increasingly governed by the magnetic field \citep{magnetosphere}. The accreted matter is funneled along the magnetic field lines directed towards the magnetic poles of the neutron star \citep{funnel_accretion}. Upon impacting the neutron-star surface, the gravitational potential energy of the accreted matter is released as X-ray radiation \citep{acrretion_to_poles}. When the magnetosphere lies close to the co-rotation radius, disk-magnetosphere interaction becomes unstable, accreted matter may accumulate near the magnetospheric boundary and be episodically released onto the neutron star surface, resulting in short-lived bursts of X-ray radiation \citep{Salvo2024}.\\
	
	These bursts can be categorized into two types based on their energy source: type-I and type-II. Generally, type-I  X-ray bursts arise from unstable thermonuclear ignition of accreted matter accumulated on the neutron star surface \citep{belian1976discovery, lewin1993xray}. Type-II bursts are attributed to instabilities in the accretion flow rather than thermonuclear burning on the neutron star surface \citep{typeII_galloway}. The recurrence time between type-II bursts ranges from a few minutes to several hours. During type-I bursts, the temperature corresponding to the blackbody emission rises due to unstable thermonuclear burning. After the burst the heated surface layer of the neutron star undergoes radiative cooling, which is shown as the decay phase \citep{jaisawal2024comprehensive}. Unlike type-I bursts that show a cooling-tail during decay phase, type-II bursts do not show a pronounced cooling phase and are characterized by nearly constant blackbody temperature throughout the temporal evolution of the burst \citep{hirotani1990interpretation}.\\

	The source IGR J17062-6143 (hereafter referred to as J17062) was discovered during the 2006 observation of \textit{INTEGRAL} \citep{Churazov2007}. The 163.65 Hz pulsations detected from the \textit{RXTE} instrument established the source as an AMXP \citep{Keek2017}. An upper limit on the magnetic field of B$\lesssim$3.5$\times$10$^8$ G was estimated by \cite{Keek2017}. \cite{vandenEijnden2018} estimated a lower limit on the magnetic field of $B\gtrsim4.8 \times 10^6$ G for low inclination angle. They also reported an observed flux of $\sim$0.98 $\times$ 10$^{-10}$ ergs cm$^{-2}$ s$^{-1}$ in the 3-79 keV energy range. \cite{Strohmayer_2018} estimated an inclination angle within 19$^\circ$ and 27.5$^{\circ}$ corresponding to a neutron star of mass 1.4M$_{\odot}$ and 2.0M$_{\odot}$, respectively. From reflection studies, a broad iron emission line has been confirmed in previous observations of J17062, showing reflection of hard X-rays from the inner accretion disk \citep{keek2017x, Fabian_2017}. Previous studies by \cite{Keek2017} showed a highly ionized disk with a low accreting flux. It is a persistent source with long-term mass accretion rate $\dot{M}\simeq2.5 \times 10^{-11} M_{\odot}$ yr$^{-1}$ \citep{keek2017x, Strohmayer_2018}. Earlier, type-I X-ray bursts have been observed from this source \citep{2013ApJ...767L..37D, keek2017x}. For instance, the burst fluences during the observed bursts in 2012 outbursts by \cite{2013ApJ...767L..37D} were in the range $(1.60-1.82)\times10^{-5}$ erg cm$^{-2}$. The average bolometric blackbody ﬂux was estimated as $\sim$4.8$\times10^{-8}$ erg cm$^{-2}$ s$^{-1}$.  \cite{keek2017x} estimated a mean blackbody temperature $kT\sim$0.29 keV with unabsorbed bolometric peak blackbody ﬂux $\sim$6.0$\times10^{-8}$ erg cm$^{-2}$ s$^{-1}$. Equating with the Eddington luminosity, the distance to the source was estimated as $d=7.3\pm0.5$ kpc. In addition, the spectral analysis suggested the reflection region moved significantly closer to the neutron star during the burst, with inner disk radius decreasing from $R_{\rm in}\sim200R_{\rm g}$ to $\sim14R_{\rm g}$. The inward migration of the disk suggests a strong influence of the burst emission on the accretion flow, leading to a modification of the inner accretion disk geometry.\\
	
	\textit{NuSTAR}$'$s unprecedented sensitivity and broad energy bandwidth, free of pileup distortions, allow us to clearly map the reflection features in various neutron star X-ray binaries. The source J17062 was effectively observed by \textit{NuSTAR} three times between 2015 and 2022. Previous \textit{NuSTAR} observations performed between 2015 and 2016 were analyzed by different authors, and broad-band spectral analysis revealed reflection features from the highly ionized disk \citep{Degenaar2017, keek2017x}. From the reflection spectroscopy, \cite{Degenaar2017} inferred a truncated disk very far from the neutron star surface at $\gtrsim$100 $R_{\rm g}$. The inferred inner radius is significantly larger than typically observed in accreting neutron stars. \cite{keek2017x} also performed spectral analysis of the 2015 \textit{NuSTAR} observation and suggested a larger inner disk radius of $\sim210 R_{\rm g}$ from the spectral fits. Later, \cite{vandenEijnden2018} analyzed the 2016 \textit{NuSTAR} observation along with the 2015 observation. They found a strongly truncated accretion disk at $\sim77 R_{\rm g}$, assuming a high inclination. They also reported that the data quality of the 2016 \textit{NuSTAR} observation was not sufficient to constrain the full broad-band relativistic reflection spectrum. However, in all such studies, some important parameters were not well constrained by the different reflection models. For example, \cite{Degenaar2017} could not constrain the inclination angle using the reflection model. Moreover, the inferred inner radius was barely constrained. Although $R_{\rm in}$ exceeded $100 R_{\rm g}$, the $R_{\rm ISCO}$ could not be excluded. \cite{vandenEijnden2018} also noted that a low inclination and a disc extending to the neutron star cannot be ruled out.\\
	
	Keeping those in mind, we performed spectral analysis of the most recent \textit{NuSTAR} observation to constrain various parameters precisely. We systematically investigated the continuum and reflection emission and applied different continuum and self-consistent relativistic reflection models to put tight constraints on the system's parameters. We detected an unusual profile of burst light curves in \textit{NICER} observations and attributed them to type-II-like bursts. We performed spectral analysis of the persistent and also the burst emission using \textit{NICER} observations five months prior to the \textit{NuSTAR} observation. This paper is structured as follows. In Section \ref{sec:obs_data}, we state the observation details and describe the data reduction process. In Section \ref{sec:lc} we discuss the properties of the light curves corresponding to the observations. In Section \ref{sec:spect_analysis}, we present spectral analysis techniques adapted for the related models corresponding to the persistent and burst emission to carry out the spectral analysis. Finally, in Section \ref{sec:disc}, we carry out the discussion based on the observations and inferences drawn from the spectral analysis.\\
	
	\section{Observation and Data reduction}
	\label{sec:obs_data}
	\textit{Nuclear Spectroscopic Telescopic Array} (\textit{NuSTAR}), launched by \textit{NASA} on 2012 June 13, is the first focusing hard X-ray telescope operating in the high energy ($3-79$ keV) range (\citealt{harrison2013nuclear}). \textit{Neutron Star Interior Composition Explorer} (\textit{NICER}) mission is dedicated towards the study of thermal and non-thermal emission from neutron stars in the 0.2-12 keV (soft) X-ray band \citep{2016SPIE.9905E..1HG}.
	The analyses presented in this work are based on observations of J17062 conducted by \textit{NuSTAR} and \textit{NICER} as tabulated in Table \ref{tab:obs_all}.
	
	\begin{table}
		\caption{Details of the observations IDs of the source J17062 that are utilized in this work.}
		\renewcommand{\arraystretch}{0.9}
		\begin{tabular}{l l c l l c c}
			\hline
			\small{Label} & \small{Observation ID}	& \small{Date (MJD)} & \small{Instrument} & \parbox{1.2cm}{\small{Exposure} \\ \small{(ks)}} &  \multicolumn{2}{c}{\parbox{2cm}{\small{Net Count Rate} \\ \small{(counts s$^{-1}$)}}} \\
			\hline
			& & & & & \small{Persistent} & \small{Peak} \\
			Obs 1 & 30801032002 & 2022.09.19 (59841.52) & \textit{NuSTAR FPMA/B} & 62.79 & 2.0 &  \\
			Obs 2 & 5034100109 & 2022.09.19 (59841.74) & \textit{NICER XTI} & 0.54 & 6 & 80 \\
			Obs 3 & 5531010105 & 2022.04.25 (59694.03) & \textit{NICER XTI} & 12.65 & 25 & 800 \\
			Obs 4 & 5531010215 & 2022.10.08 (59860.31) & \textit{NICER XTI} & 1.4 & 25 & 465 \\
			\hline
			\renewcommand{\arraystretch}{1.0}
		\end{tabular}
		\label{tab:obs_all}
	\end{table}
	
	The \textit{NuSTAR} data were processed using the data analysis software \texttt{NuSTARDAS v0.4.12}, within \texttt{HEASOFT v6.36} employing the latest calibration database (\texttt{CALDB v20260421}). The task \texttt{nupipeline v0.4.12} was used to generate the calibrated and screened event files. For all the observations a circular extraction region of $120''$ radius was used to study the persistent and burst spectrum. From the same chip, the background was selected as a region of radius $120''$ far away from the source. HEASOFT provides FTOOLS to deduce the $FITS$ files needed for analysing the observed data. The FTOOL \texttt{nuproducts} was used to extract the spectra and light curves from FPMA and FPMB detectors. The \textit{NICER} standard calibration, screening, and filtering of the data is done using the \texttt{nicerl2 v1.41} pipeline \citep{remillard2022empirical}. The lightcurve were produced using \texttt{nicerl3-lc} task. The screened files obtained from \texttt{nicerl2} pipeline are further utilised to extract the required files for spectral analysis using \texttt{nicerl3-spect} task.
	
	
	For studying the bursts, good time interval (GTI) files were generated separately for the persistent emission and the observed type-II-like X-ray burst(s) by using the \texttt{maketime} task on the \textsc{mkf} file located in \textit{auxil} directory of the corresponding \textit{NICER} observation. During burst analysis, data above 5.0 keV were excluded due to background dominance. The details of the observation IDs included in this paper are provided in Table \ref{tab:obs_all}. Hereafter, the \textit{NuSTAR} observation ID 30801032002, and \textit{NICER} observation IDs 5034100109, 5531010105, and 5531010215 are referred to as Obs 1, 2, 3, and 4, respectively.
	
	\section{Light Curve}
	\label{sec:lc}
	
	\begin{figure}
		\centering	
		\includegraphics[width=0.45\columnwidth]{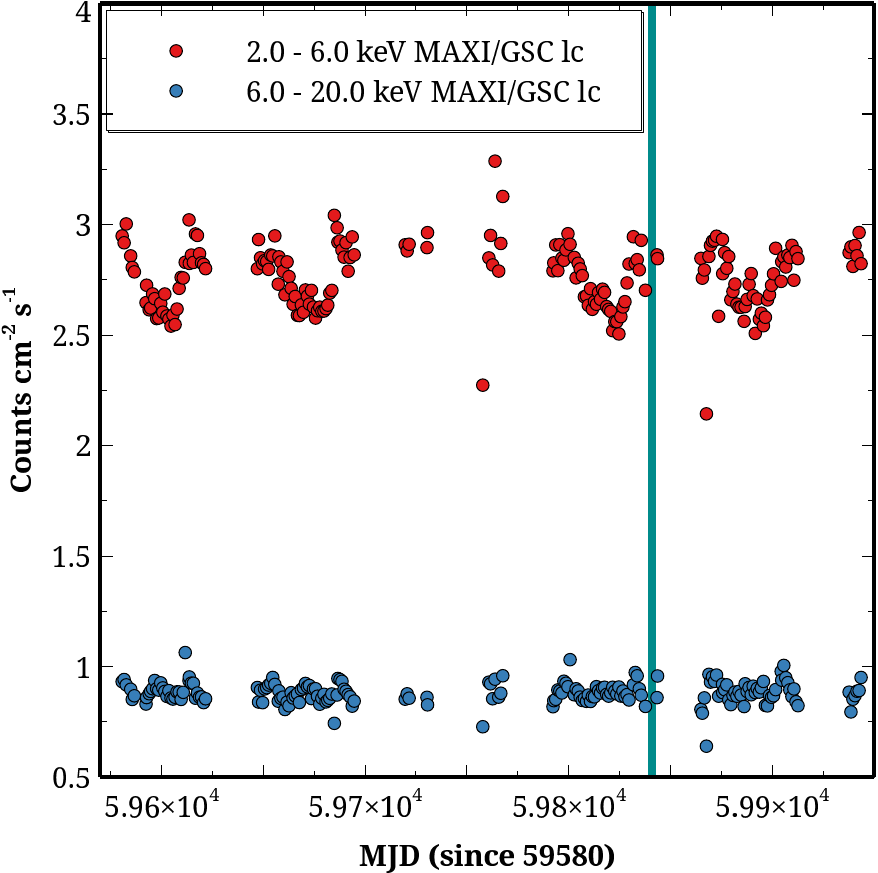}
		\includegraphics[width=0.45\columnwidth]{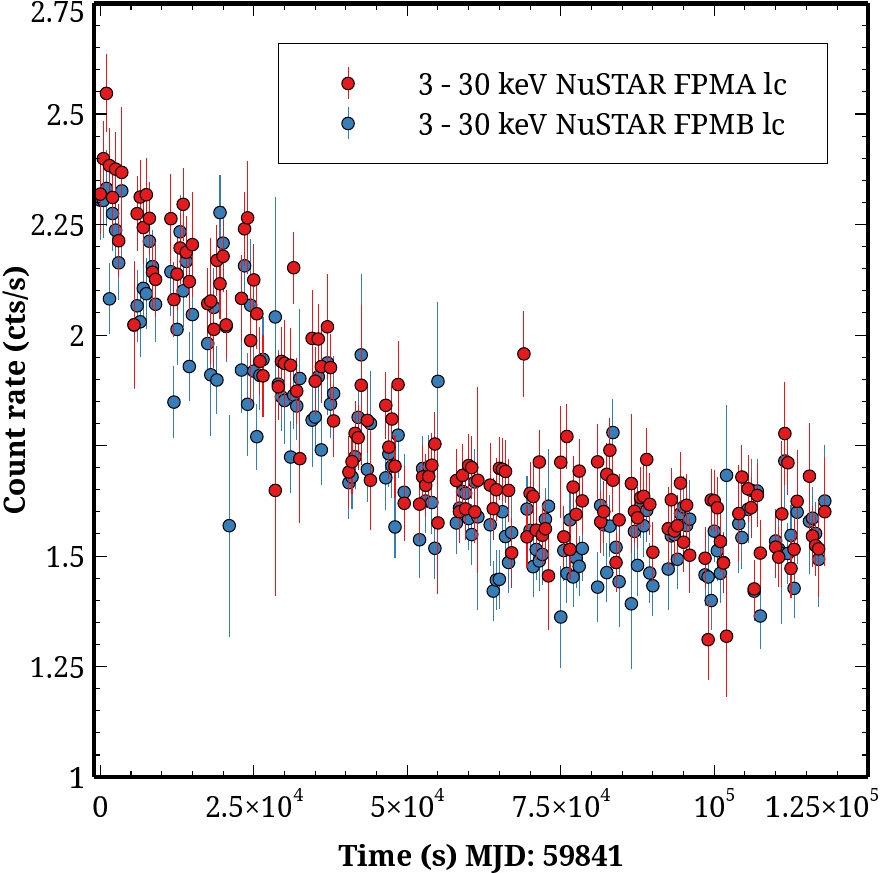}
		\includegraphics[width=0.45\columnwidth]{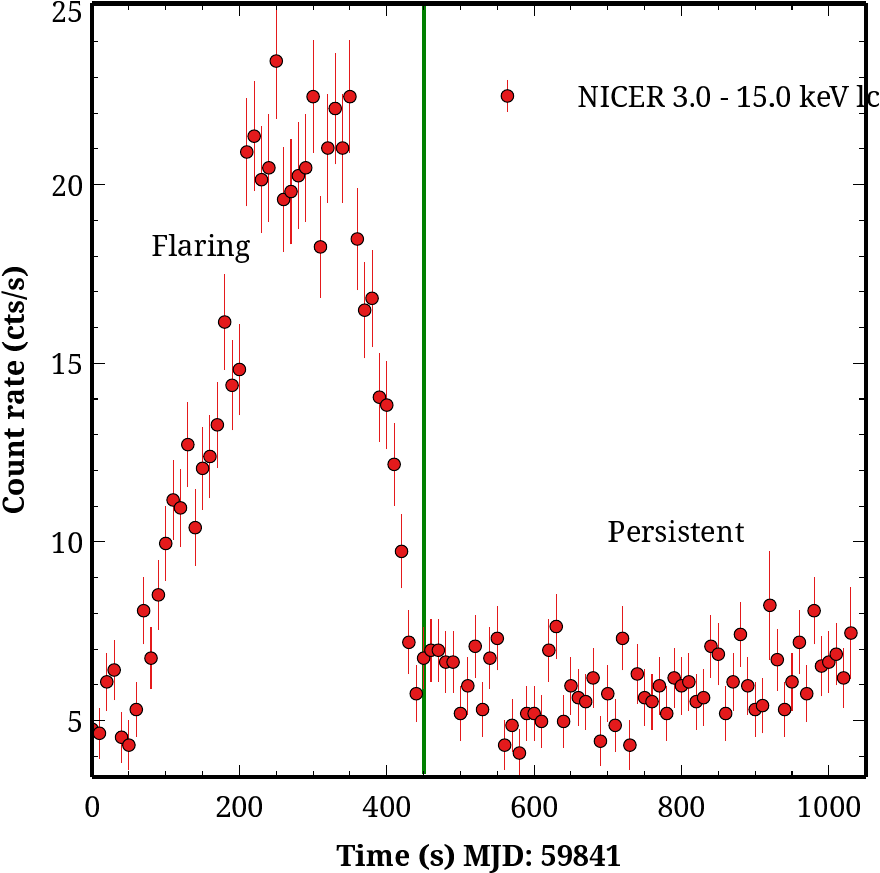}
		\includegraphics[width=0.45\columnwidth]{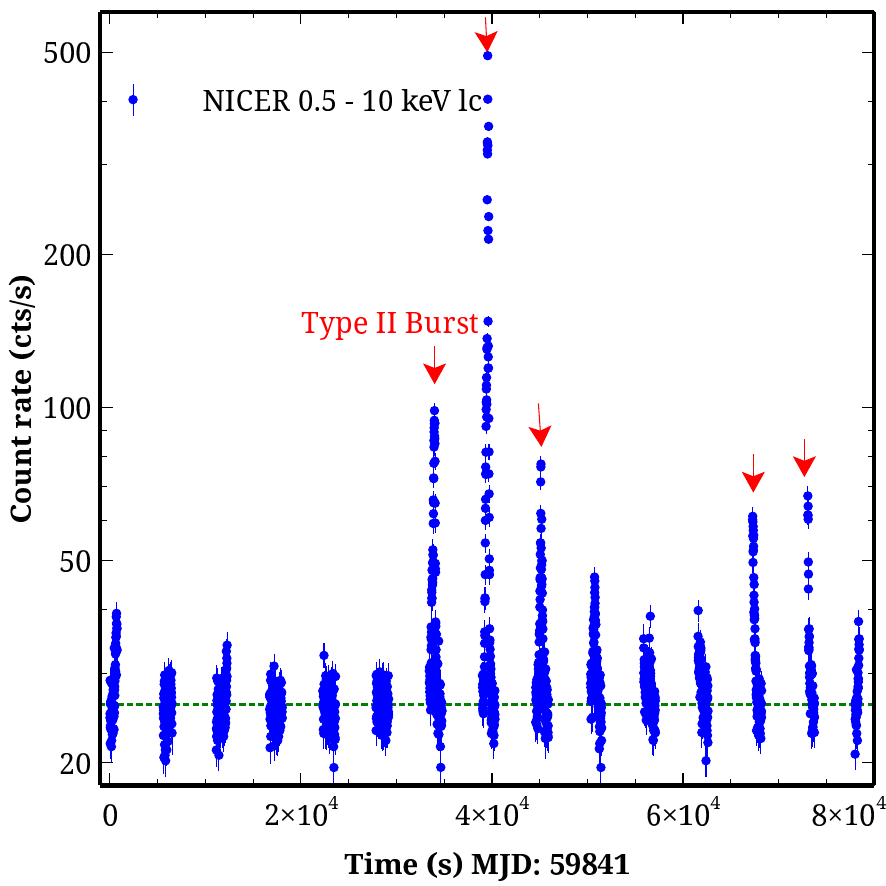}
		\caption{Light curve of the source J17062 for -- \textit{upper left panel:}  \textit{MAXI} observation in 2-6 keV energy range (\textit{red}) and 6-20 keV energy range (\textit{blue}) for the long-term exposure of 420 days from 2022 January 01 (MJD $59580$) to 2023 February 25 (MJD $60000$), wherein Obs 1 and 2 (Table \ref{tab:obs_all}) performed on 2022 September 19 are shown by the \textit{sea green} colored line, \textit{upper right panel:} \textit{NuSTAR} observation (Obs 1) in 3-30 keV energy range with a 10 s bin size corresponding to the instruments FPMA(\textit{red}) and FPMB(\textit{blue}), \textit{lower left panel:} \textit{NICER} observation (Obs 2) where the vertical green line demarcates the two regions corresponding to the flaring (\textit{left}) and persistent (\textit{right}) emission in the 3-15 keV energy range with a 10 s time bin, and \textit{lower right panel:} \textit{NICER} observation (Obs 3) in 0.5-10.0 keV energy range with a time bin of 10 s, where the red arrows indicate the observed type-II-like bursts and the dashed green line indicates the persistent level.}
		\label{fig:all_lc}
	\end{figure}
	
	\begin{figure}
		\centering
		\includegraphics[width=0.32\columnwidth]{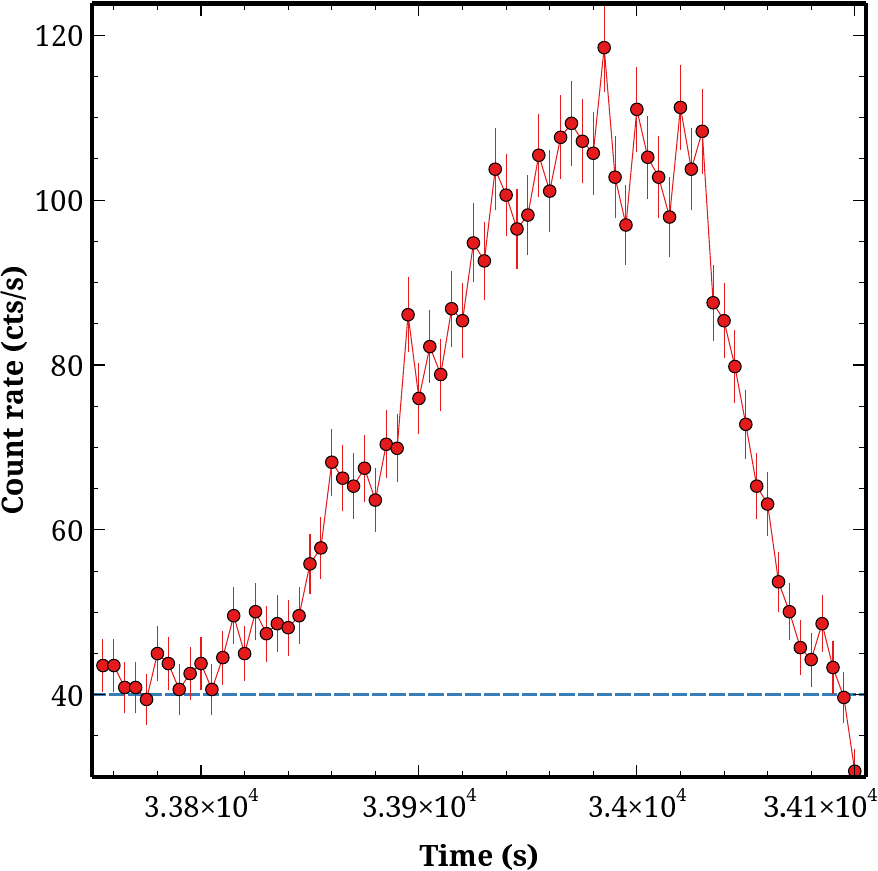}
		\includegraphics[width=0.32\columnwidth]{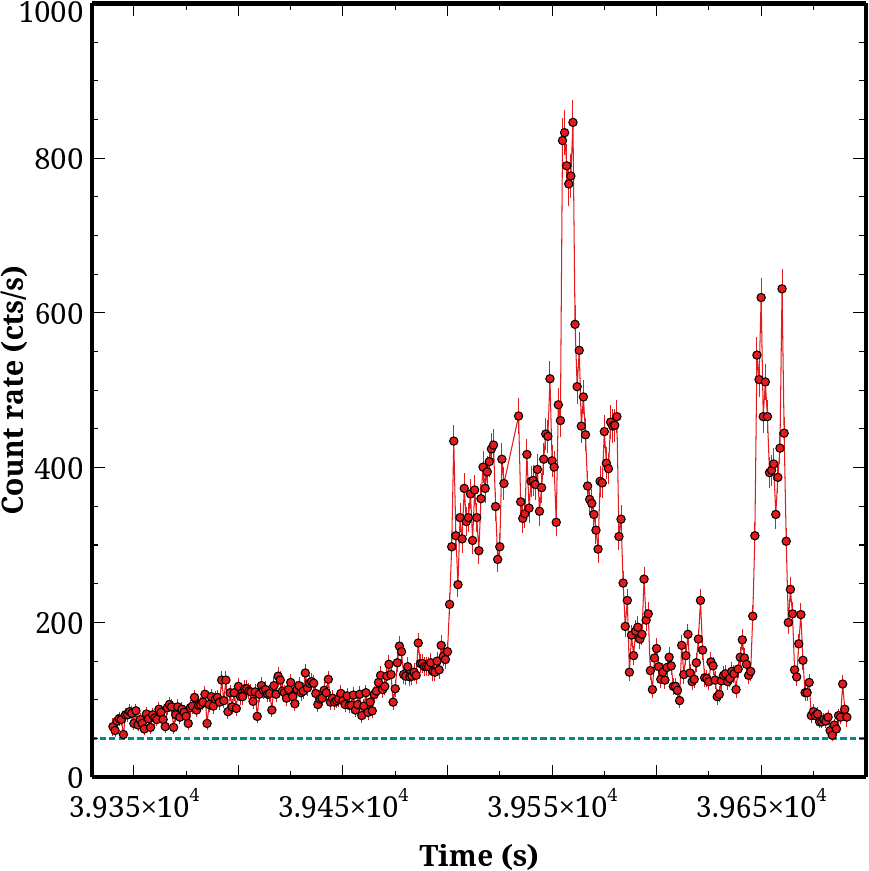}
		\includegraphics[width=0.32\columnwidth]{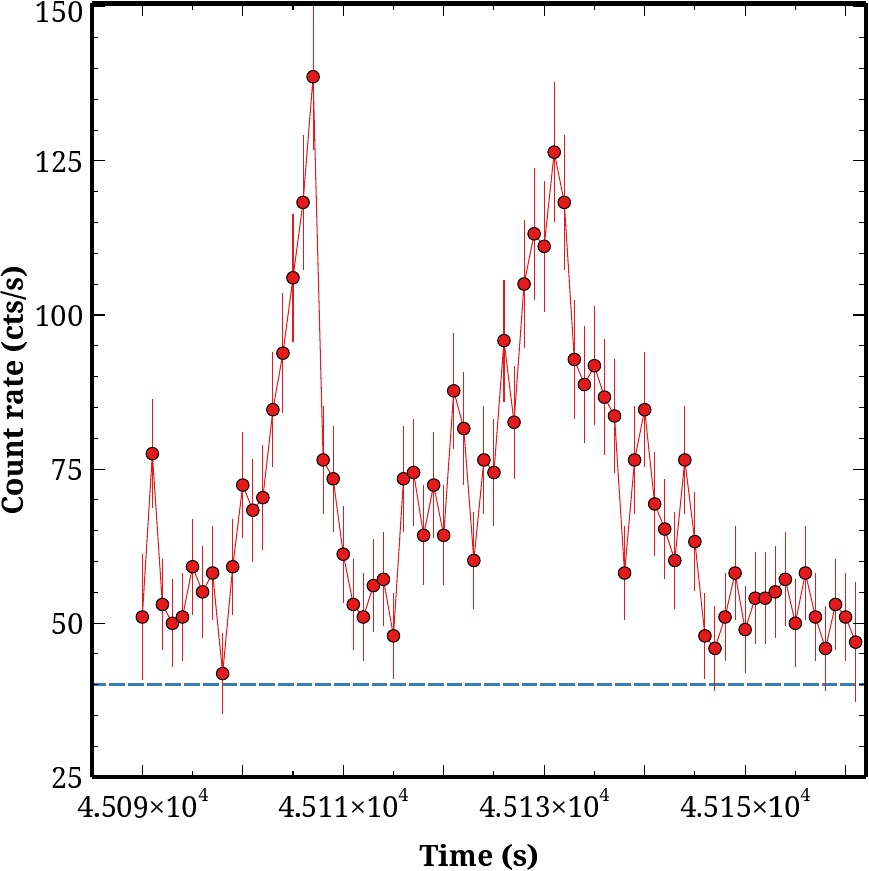}
		\includegraphics[width=0.45\columnwidth]{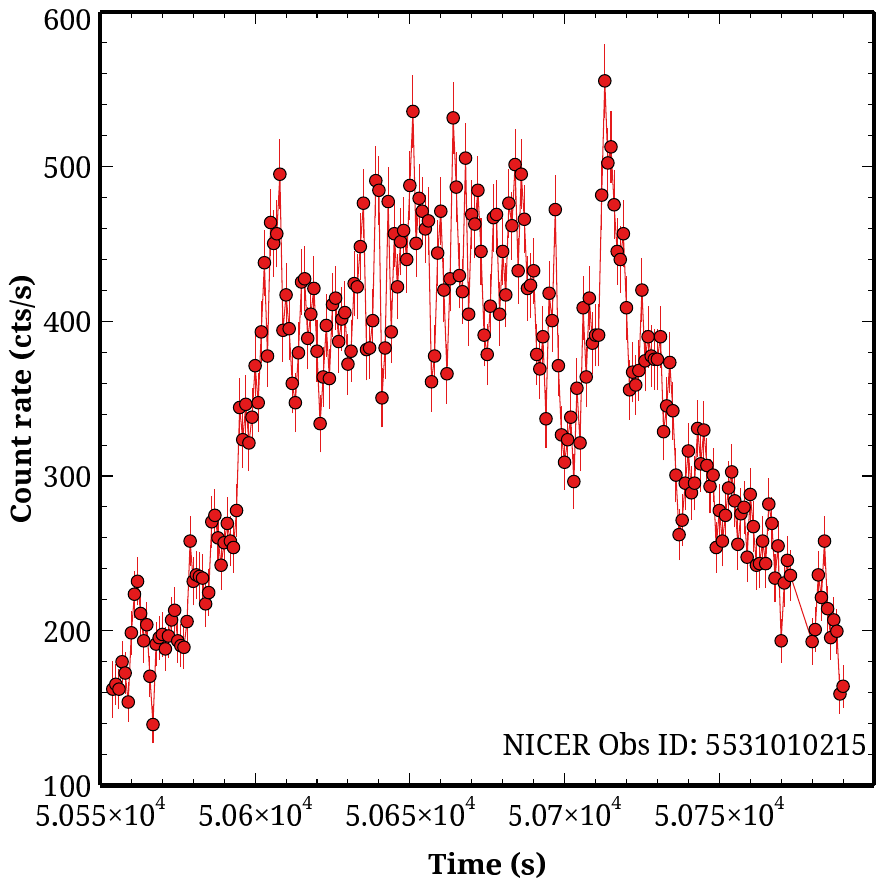}
		\caption{Light curves of the source J17062 for -- Obs 3 corresponding to the bursting regions where the emission increases above the persistent emission - \textit{upper left panel:} first burst (B1) with a time bin of 10 s, \textit{upper middle panel:} second burst (B2) with a 5 s time bin and \textit{upper right panel:} third burst (B3) with a time bin of 10 s -- in the 1-15 keV energy range, - and \textit{bottom panel:} Obs 4 in the 1-15 keV energy regime with a time bin of 10 s. The blue dashed line in each plot on the top panel shows the persistent level for that particular \textit{NICER} observation.}
		\label{fig:bursts}
	\end{figure}

	The long-term \textit{MAXI}/GSC light curve of the source J17062 is shown in the \textit{upper left} panel of Figure \ref{fig:all_lc}, indicating the \textit{NuSTAR} FPMA/B observation considered for the present analysis \citep{2009PASJ...61..999M}. The complete \textit{NuSTAR} FPMA/FPMB and \textit{NICER}/XRT light curve corresponding to Obs 1, 2, and 3 are shown in the \textit{upper right}, \textit{lower left}, and \textit{lower right} panel of Fig. \ref{fig:all_lc}. \textit{NuSTAR} observation (Obs 1) presents an overall persistent emission having a count rate within $\sim$ 1.5-2.5 counts s$^{-1}$. For \textit{NICER} observation (Obs 2), the count rate during persistent emission varies within $\sim$4-7 counts s$^{-1}$. We observe a region during the initial exposure of Obs 2, wherein the count rate reaches 22 counts s$^{-1}$ within $\sim$200 s. The peak emission stalls for $\sim$150 s duration before falling sharply back to 5 counts s$^{-1}$ within 35 s. Although there is a $\sim$5 times increase in the net count rate, the event does not adhere to a burst profile. We are now making a difference with burst regions. Essentially, we refer to such intervals where the emission rises to a peak value within a certain duration and then decays to its prior level as a flaring region. For the Obs 2, 3, and 4, the net count rates corresponding to the persistent emission and peak regions -- depending on the presence of flaring or bursting intervals within the observations -- are listed in Table \ref{tab:obs_all}. For Obs 3 and 4, the peak emissions correspond to the bursts with the maximum count rate. The net count rate for the persistent emission for Obs 3 varies within $\sim$10-30 counts s$^{-1}$. For Obs 4, the persistent emission remains within $\sim$10-40 counts s$^{-1}$. Apart from light curve studies, we have not considered Obs 2 and 4 further. We mainly focus on Obs 3.\\
	
	From the light curve of the Obs 3, as shown in the \textit{lower right} panel of Fig. \ref{fig:all_lc}, five prominent flaring regions are observed. These flares show burst-like features and are identified as five distinct burst events, as indicated by small red arrows in the light curve. Out of the five observed bursts, we have considered the first three (\textit{viz.} B1, B2, and B3), for which the full rise and decay regions were captured during the \textit{seventh}, \textit{eighth}, and \textit{ninth} \textit{NICER} cycle of the particular exposure.  The light curves corresponding to the rise, peak, and decay of the bursts are plotted, with a horizontal line indicating the persistent emission level, and are shown in the \textit{upper} panels of Fig. \ref{fig:bursts}. The peak count rate during B1, B2, and B3 are $\sim$113, $\sim$843, and $\sim$138 counts s$^{-1}$, respectively. We observe that the burst profiles do not clearly indicate the characteristics of fast rise and exponential decay (FRED), which are commonly observed in type-I bursts. We therefore attributed the observed bursts as type-II-like bursts. The light curve profiles of the typical type-II bursts differ from those of type-I bursts in a number of features. Type-II bursts exhibit a variety of morphological modes that may suddenly switch between them. Type-II bursts are the most energetic, longest, and least frequent, commonly associated with a hard spectral state when the persistent luminosity is below 10 percent of the Eddington limit. \\
	
	In our case, we have observed type-II-like bursts with different morphologies/patterns. The first burst of \textit{NICER} Obs 3 (B1) is about $\sim$350 seconds long and has a flat peak lasting at least $\sim$100 seconds. The burst observed in \textit{NICER} Obs 4 has similar characteristics: a long duration (around $\sim$250 seconds), a high intensity, and a long plateau. These features may be associated with the so called mode-0 type-II bursts \citep{typeII_galloway, type2_2026}. However, the characteristics of the second (B2) and third (B3) bursts observed from \textit{NICER} Obs 3 are completely different. For bursts B2 and B3, we observed the presence of multiple peaks associated with individual events. These features are not strictly associated with mode--1 type-II bursts, which have a characteristic sequence of rapidly recurring bursts. Rather, they may resemble the switching of mode--0 to mode--1 of type-II bursts. Moreover, after the first long mode-0-like type-II burst (B1), we observed that the persistent emission enhances to a level higher than that of the intraburst emission (see the beginning of B2). This is the so-called `hump' shape observed during the switching of mode--0 to mode--1 type-II burst. There is another pattern of type-II bursts, known as mode--2, which has not been observed in our case.\\
	
	Apart from the above-mentioned observations, there are some other \textit{NICER} observations \textit{viz.} observation IDs 5531010212, 5531010214, 5531010215, and 5531010216, that also display type-II characteristics. The \textit{NICER} observation 5531010215 (Obs 4) shows a longer peak burst phase ($>100s$) compared to the rising phase ($\sim$100s) of the burst. We will consider all the type-II-like bursts observed by \textit{NICER} in our future work.  Presently, we undertook spectral analysis of the first three bursts of Obs 3 for which complete burst phases were present. For the remaining two bursts B4 and B5 of Obs 3, the peak count rates were $\sim$92 and $\sim$102 counts s$^{-1}$, respectively. Their rising and peak phases were occulted, and hence were not included for burst spectral analysis. \\
	
	There is a significant time lapse during the rising phase of each of the bursts B1, B2 and B3. In our analysis, we considered the burst initiation level as the time where the count rate exceeds 50$\%$ of the persistent net count rate. The burst rise time is considered as the time taken to reach 90$\%$ of the peak burst count rate \citep{galloway2008thermonuclear}. Time duration for peak burst is the duration for which the emission remains above 90$\%$ of the peak level. Time taken for the emission to go below the burst initiation level is considered the decay time. For B2 and B3, another lower maximum is observed after the peak burst. The details for the bursts of Obs 3 are listed in Table \ref{tab:Burst_times}.\\

	\begin{table*}
		\centering
		\caption{Time interval of the bursts with the rise, peak and decay time for the observed bursts B1, B2, and B3 for the \textit{NICER} Obs 3 of the source J17062 in the energy regime 1-15 keV. For each of the bursts, duration of burst exposures are noted with the individual peak counts.}
		\renewcommand{\arraystretch}{0.9}
		\begin{tabular}{llllccr}
			\hline
			\parbox{1cm}{\small{Burst} \\ \small{Serial} \\ \small{no.}} & \parbox{4.2cm}{\small{Start time(s) - End time(s)}} & \parbox{1cm}{\small{Burst} \\ \small{exposure} \\ \small{(s)}} & \parbox{0.6cm}{\small{t$_{rise}$} \\ \small{(s)}} & \parbox{0.5cm}{\small{t$_{peak}$} \\ \small{(s)}} & \parbox{0.5cm}{\small{t$_{decay}$} \\ \small{(s)}}  & \parbox{1.5cm}{\small{Peak} \\ \small{intensity} \\ \small{(counts s$^{-1}$)}} \\
			\hline
			B1 & 33749 - 34100 & 351 & 188 & 83 & 45 & 126 \\
			& & & & & & \\
			B2 & 39320 - 39720 & 350 & \parbox{1cm}{235 \\ 12} & \parbox{1cm}{\, \, \, 6$^{\dagger}$ \\\, \, 2$^{\dagger\dagger}$} & \parbox{1cm}{ \, \, 64 \\ \, \, \,  31} & \parbox{1cm}{\, \, 843$^\dagger$ \\ \, \, \, 621$^{\dagger\dagger}$} \\
			& & & & & &\\
			B3 & 45090 - 45160 & 70 & \parbox{1cm}{16 \\ 15} & \parbox{1cm}{\, \, 2$^{\dagger}$ \\ \, \, 3$^{\dagger\dagger}$} & \parbox{1cm}{\, \, 7 \\ \, \, \, 27} & \parbox{1cm}{\, \, 138$^{\dagger}$ \\ \, \, \, 126$^{\dagger\dagger}$} \\
			B4 & 67310 - 67790 &  480 & & 127 & 353 & 65 \\
			B5 & 73050 - 73340 &  290 & & 50 & 240 & 75 \\
			\hline
			\textit{Note:} & \multicolumn{6}{l}{$^{\dagger}$ denotes the emission for 1$^{st}$ peak; $^{\dagger\dagger}$ denotes the emission for 2$^{nd}$ peak.}
			\renewcommand{\arraystretch}{1.0}
		\end{tabular}
		\label{tab:Burst_times}
	\end{table*}

	\section{Spectral Analysis}
	\label{sec:spect_analysis}
	\subsection{\textit{NuSTAR}}
	\subsubsection{Persistent emission}
\begin{figure}
		\centering
		\includegraphics[height=6cm, width=0.45\columnwidth]{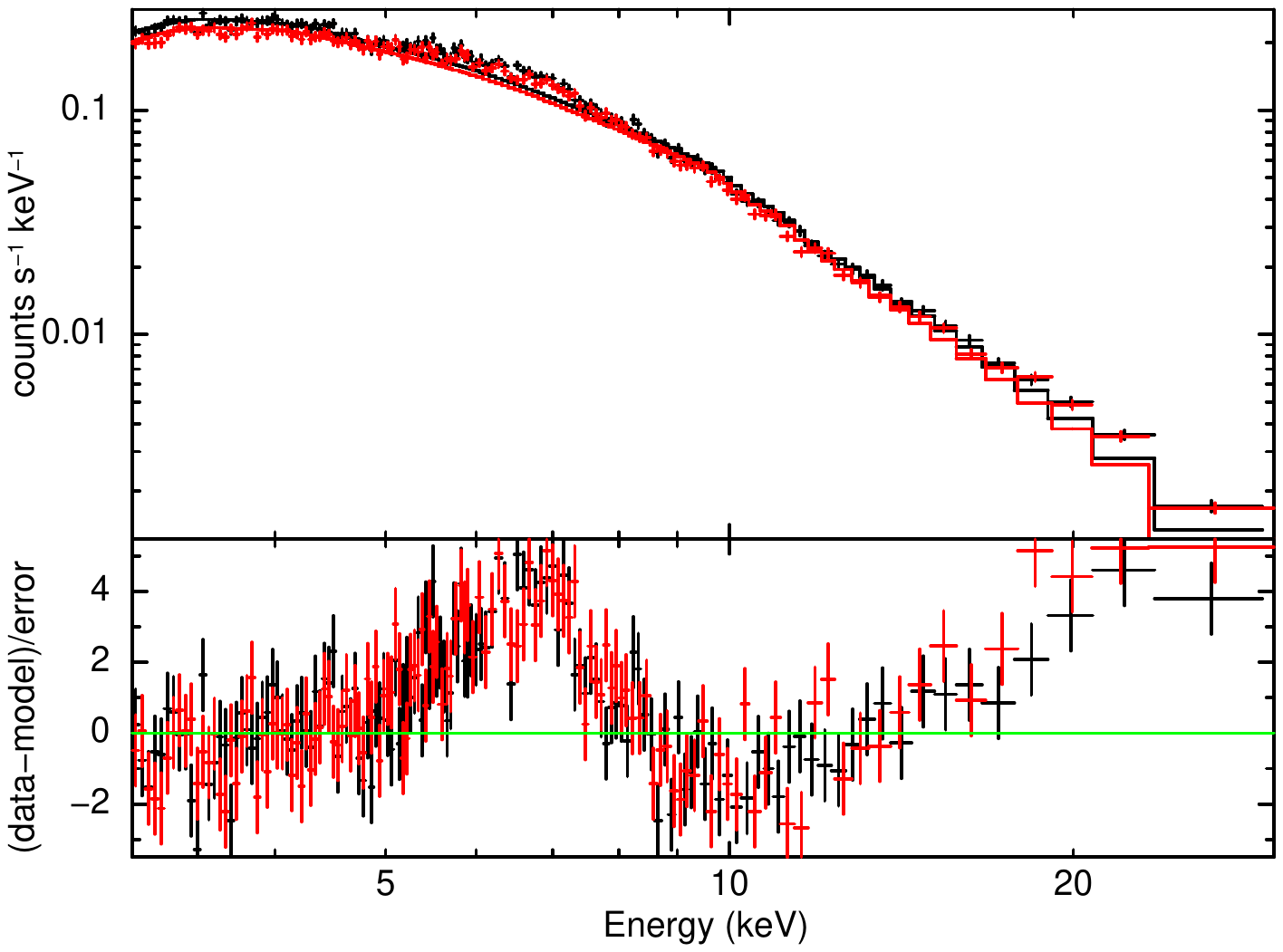}
		\includegraphics[height=6cm, width=0.45\columnwidth]{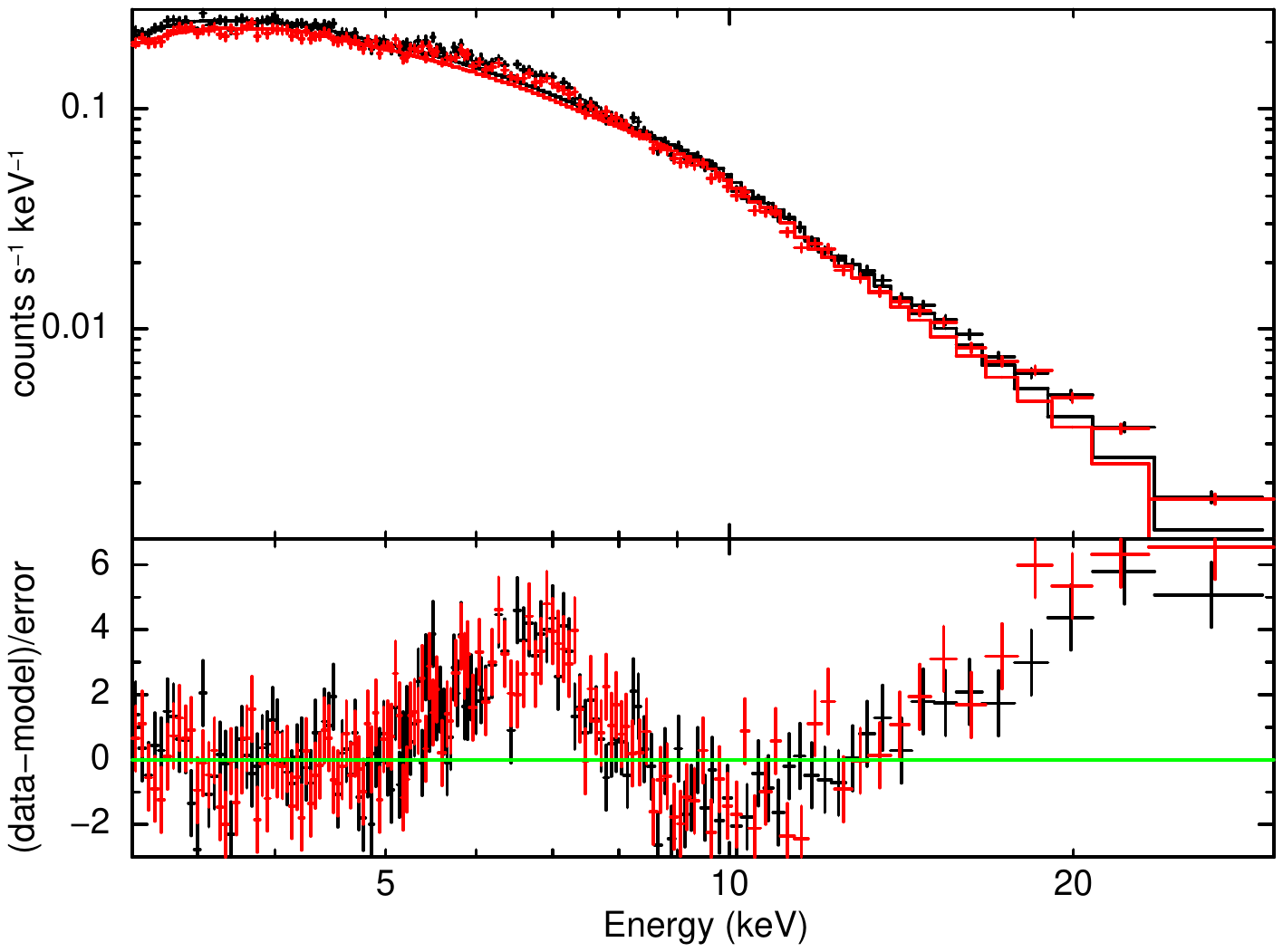}
		\caption{Unfolded spectral data for the best-fit corresponding to the \textit{left panel:} model \texttt{const*TBabs*powerlaw}, and \textit{right panel:} model \texttt{const*TBabs*(nthcomp+diskbb)} is shown for the \textit{NuSTAR} spectrum (Obs 1) of FPMA (black) and FPMB (red) of the source J17602 in the energy range 3-30 keV. Lower panels within each of the panel show the residuals with 1$\sigma$ error bar.}
		\label{fig:per_bursts}
	\end{figure}

	The spectral analysis for the persistent and burst emission was carried out using \texttt{XSPEC v}12.15.1 (\citealt{1996ASPC..101...17A}). To account for the interstellar absorption along the line of sight, we have used the \texttt{TBabs} model. Initially, we tried to fit the 3-30 keV \textit{NuSTAR} FPMA and FPMB spectra simultaneously with an absorbed power-law model. It provided a good description of the continuum emission with $\chi^2/dof$ = 1177/542 and a power-law photon index, $\Gamma$, of around $\sim$2.5. The best-fit spectral plot is shown in the \textit{left} panel of Figure \ref{fig:per_bursts}, along with the fit residuals. We then introduced the physically motivated model \texttt{nthcomp} to fit the continuum emission instead of a power-law (\citealt{1996MNRAS.283..193Z}, \citealt{1999MNRAS.309..561Z}). It is a thermal Comptonization model, which involves a seed photon source as a blackbody or a disk-like multi-color blackbody. The seed photons are up-scattered by hot electrons, resulting in the output spectrum. We considered a geometry in which the seed photons are arising from the disk. It provided $\chi^2/dof$=1081/540  with photon index ($\Gamma$) and disk temperature ($kT_{\rm in}$) of around $\sim$2.54 and $\sim$0.75 keV, respectively. However, the fit revealed a high electron temperature ($kT_{\rm e}$) of around $\sim$550 keV. The corresponding spectral plot is shown in the \textit{right} panel of Figure \ref{fig:per_bursts}. \\
	
	We further replaced the \texttt{nthcomp} with its updated version, \texttt{thcomp} convolved with \texttt{diskbb}. \texttt{Thcomp} describes the continuum Comptonization emission much better than a cutoff power law model, and it aligns more closely with the actual Monte Carlo spectra from Comptonization than \texttt{nthcomp} does. The \texttt{thcomp} model estimates the Comptonization distribution of seed photons from the disk, that are upscattered by the corona, as well as, the neutron star surface simultaneously \citep{1996MNRAS.283..193Z, 2019MNRAS.485.2942N, 2020MNRAS.492.5234Z}. It further improved the fit with $\chi^2/dof$=860/539$\sim$1.60. The electron temperature reached the upper hard limit provided by the models, namely 1000 keV for \texttt{nthcomp} and 1500 keV for \texttt{thcomp*diskbb}, respectively. The parameters for \texttt{thcomp*diskbb} components, photon index ($\Gamma_{\tau}$), multicolor disk temperature ($kT_{\rm in}$), electron temperature ($kT_{\rm e}$) and covering fraction ($f_{\rm cov}$) are estimated around $\sim 1.90$, $\sim1.41$ keV, $\gtrsim 100$ keV and $\sim 0.34$, respectively. Our continuum modeling showed no evidence for an observable high-energy cutoff ($E_{\rm cut}$) in the \textit{NuSTAR} bandpass. Since the high-energy cutoff is related to the electron temperature of the corona, the electrons in the plasma must be very hot ($kT_{\rm e} \gtrsim$100 keV), as reflected in the fits. A high $E_{\rm cut}$ value is also evident from an earlier \textit{NuSTAR} observation in 2015 \citep{Degenaar2017, Keek2017}. We note that, despite improved statistics, the disk temperature ($kT_{\rm in}$) obtained from the model \texttt{thcomb*diskbb} is quite high and not in agreement with previous studies of this source \citep{keek2017x, vandenEijnden2018}. Moreover, the model estimated an unphysical value of the normalization parameter. Additionally, all choices of the continuum models exhibit strong residuals around 5-8 keV and 20-30 keV, signatures of the broad Fe emission line and Compton hump, respectively. It shows the robustness of the detection of the reflection features from the \textit{NuSTAR} spectrum. Detection of these reflection features in the spectrum (see Fig. \ref{fig:per_bursts}) motivates further use of self-consistent reflection models to describe them. 
	
	\subsubsection{Reflection spectroscopy}
	The hard X-ray emission produced by Comptonization in the corona irradiates the accretion disk and is reprocessed, giving rise to a reflection spectrum. The reflected Comptonized emission provides a probe into the physical properties and geometry of the accretion disk. To model the reflection spectrum, we used the relativistic reflection model \texttt{relxill} \citep{relxill}. It describes the reflection of a power-law continuum, originating from the corona, from the accretion disk. The model parameters corresponding to the \texttt{relxill} component are inclination angle of the source to the line of sight ($i$), inner disk radius ($R_{\rm in}$), outer disk radius ($R_{\rm out}$), breaking radius ($R_{\rm br}$), emissivity index $q_1$ between $R_{\rm in}$ and $R_{\rm {out}}$ and $q_2$ between $R_{\rm out}$ and $R_{\rm {br}}$, spin parameter ($a^{*}$), photon index ($\Gamma$), ionization of the accretion disk (log $\xi$), iron abundance ($A_{\rm Fe}$), high energy cut-off of the observed spectrum ($E_{\rm {cut}}$), reflection fraction, redshift ($z$) and normalization.  We also used the \texttt{relxillCP} model, which is a self-consistent reflection model. It provides a physically motivated description of the disk-reflected Comptonized coronal-emission. Except $E_{\rm {cut}}$, all other parameters of \texttt{relxill}, along with the two parameters, \textit{viz.}, density of the accretion disk (log $N$)and electron temperature ($kT_{\rm e}$), are estimated by the \texttt{relxillCP} model.\\
	
\begin{figure}
		\centering
		\includegraphics[width=0.45\columnwidth]{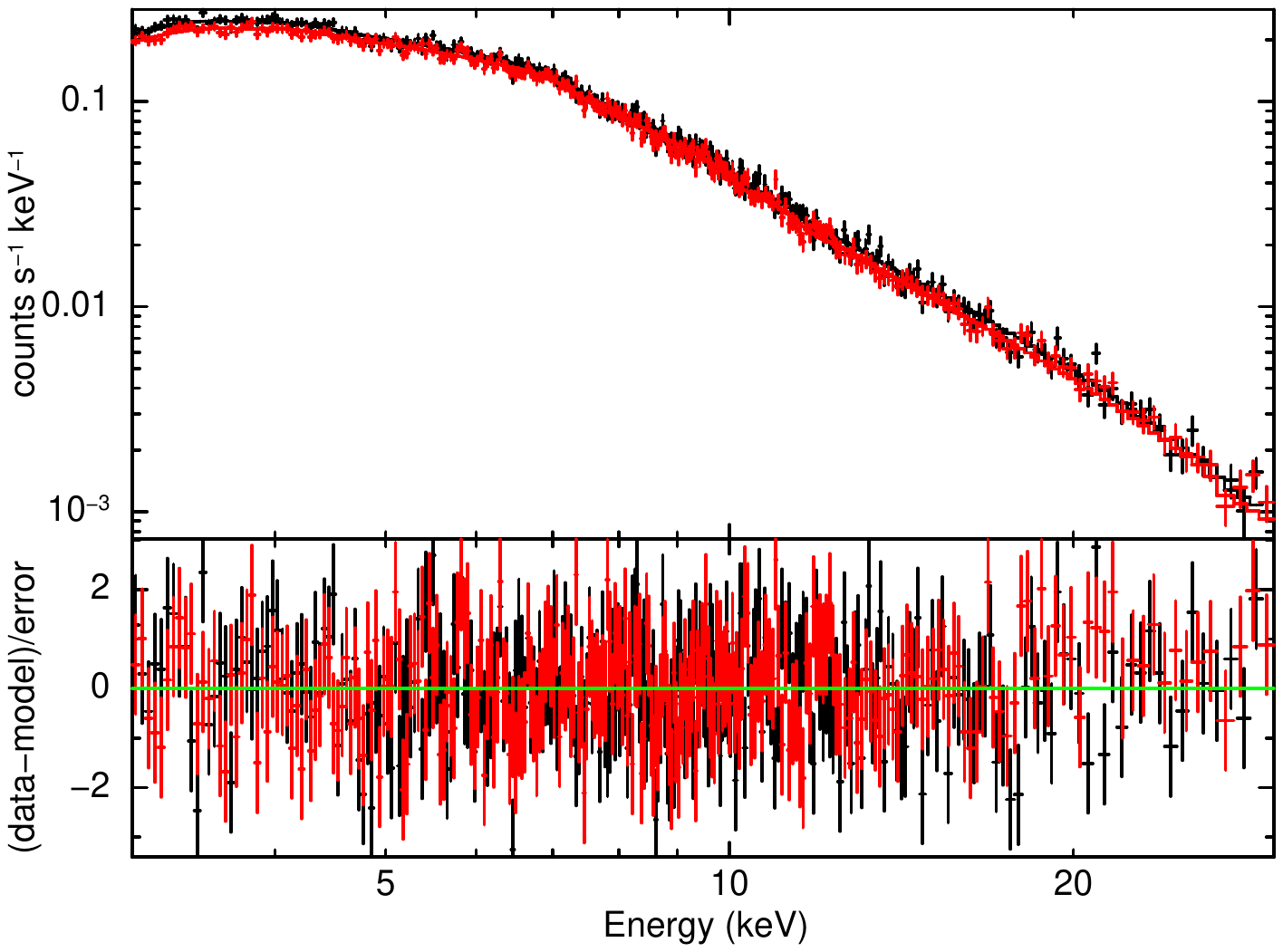}
		\includegraphics[width=0.45\columnwidth]{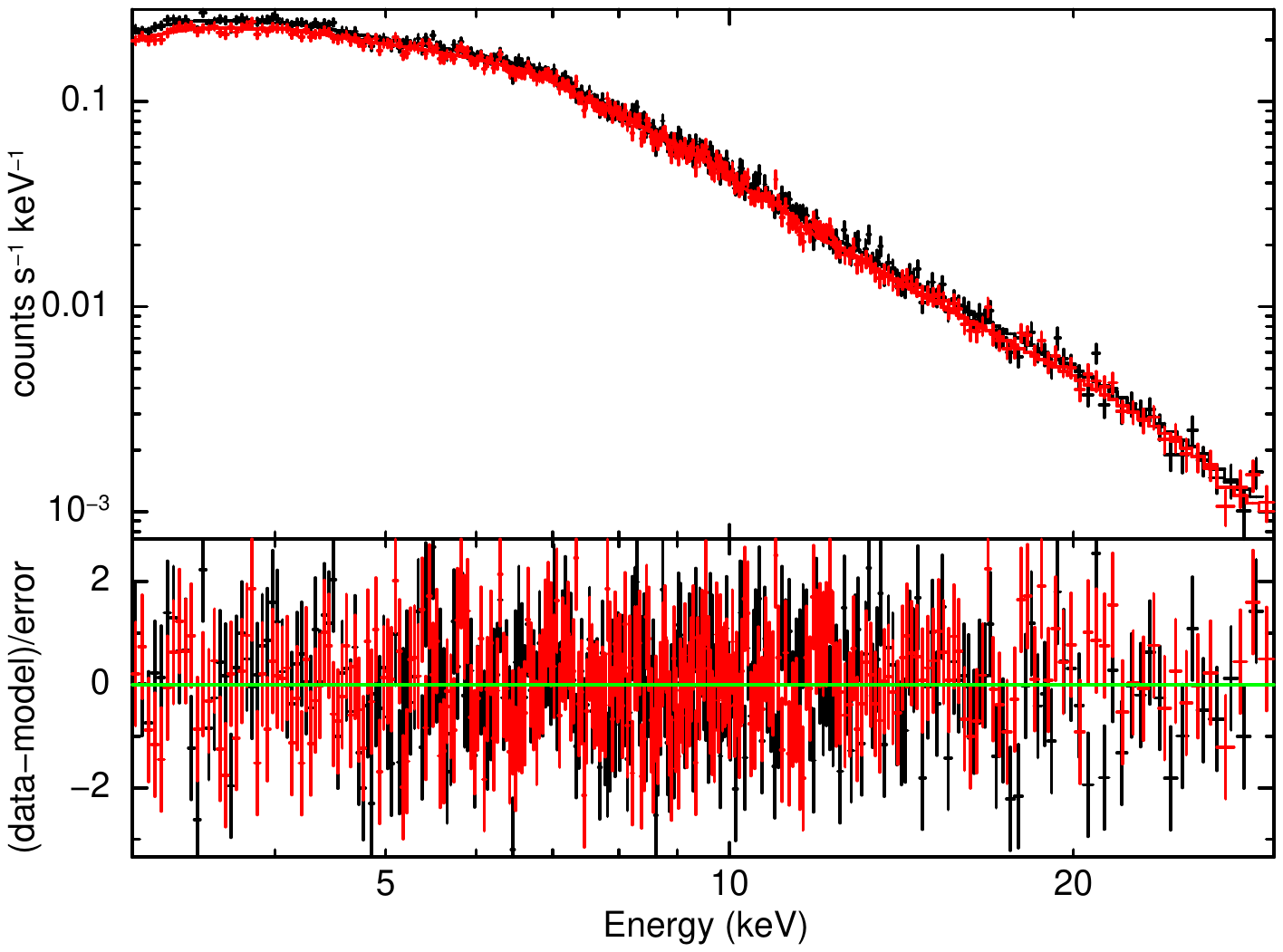}
		\caption{Unfolded spectra used for reflection modeling showing the best-fit data for the \textit{left panel:} Model 1a \texttt{const*TBabs*relxill}, and, \textit{right panel:} Model 1b \texttt{const*TBabs*relxillCP} for \textit{NuSTAR} observation (Obs 1) of FPMA (\textit{black}) and FPMB (\textit{red}) in the energy range 3-30 keV for the source J17062. The lower panels of each plot show the residuals with 1 $\sigma$ error bar.}
		\label{fig:relxill}
	\end{figure}

	\sloppy{We used the two models,  Model 1a -- \texttt{const*TBabs*relxill}, and Model 1b -- \texttt{const*TBabs*relxillCP}, to fit the persistent and reflection emission and thereby ascertain the inner radius of the accretion disk ($R_{\rm in}$) and the angle of inclination of the source ($i$). During the analysis, the neutral hydrogen column depth density ($N_{\rm H}$) is kept constant at $\sim$ $0.1 \times 10^{22}$ cm$^{-2}$ \citep{1985SSRv...40..287V, 1990ARA&A..28..215D}. The outer disk radius was frozen at 1000$R_{\rm g}$. The emissivity indices ($q_1$ and $q_2$) were frozen at 3, corresponding to the standard value of a centrally radiating source. The dimensionless spin parameter ($a^{*}$) was calculated numerically utilizing the maximum spin parameter for neutron stars in LMXBs ($j_{\rm {max}}$ $\sim$ 0.7; \citealt{lo2011spin}) and the pulsation frequency ($f$) of J17062 ($f=163.65$ Hz). The maximum breakup frequency ($f_{\rm k}$) is considered within 1000-1500 Hz. We considered the breakup frequency $\sim$1400 Hz. For slowly rotating stars ($a^* < 0.3$), the spin parameter ($a^{*}$) scales as $a^{*}=j_{\rm max}\frac{f}{f_{\rm k}}$. The value obtained is $a^{*} \simeq 0.1$, which was frozen during the spectral fitting. The redshift was fixed at $z=0$, as the emission and absorption processes were modeled in the rest frame of the source. For the Model 1a and Model 1b, we obtained best-fit statistics $\chi^2/dof$ as $516/536\sim0.96$ and $502/535\sim0.94$, respectively. The best-fit values corresponding to models 1a and 1b are listed in Table \ref{tab:pers_nustar}. The constant parameter that ensures calibration among \textit{FPMA} and \textit{FPMB} resulted in $\sim0.99$ for both models. The upper limit of the inner disk radius and inclination angle were estimated at $3.60 R_{\rm ISCO}$ and $2.60 R_{\rm ISCO}$; and  ${\sim42^{\circ}}$ and ${\sim35^{\circ}}$, respectively, for \texttt{relxill} and \texttt{relxillCP} models at a 90\% confidence level. Both reflection models indicate relatively low values of the inner disc radius ($R_{\rm in}$) and the inclination angle ($i$). The obtained photon index values for the two models are comparable. We obtained a high disk ionization parameter of $\log \xi$ around $\sim$3.4--3.8, which is consistent with the earlier estimation \citep{Degenaar2017, keek2017x}. However, Model 1a and Model 1b provided substantially different solutions for some other relevant reflection-related parameters. In particular, the iron abundance ($A_{\rm Fe}$) and the reflection fraction change significantly when \texttt{relxill} is replaced by \texttt{relxillCP}. $A_{\rm Fe}$ changes from $\sim$1.9 to $\sim$4.9, close to its upper boundary, while reflection fraction changes from $\sim$1.11 to $\sim$0.19, when the reflection model changes from \texttt{relxill} to \texttt{relxillCP}. Moreover, the \texttt{relxillCP} model estimates a higher disk density ($\log N$) of around $\sim$19.1 (in units of cm$^{-3}$), which is also close to the upper boundary of the adopted model grid. As we obtained high electron temperature corresponding to the continuum models, in a similar manner, for Model 1a (\texttt{relxill}) the cutoff energy ($E_{\rm cut}$) provided a minimum value $\sim$930 keV, and for Model 1b (\texttt{relxillCP}), electron temperature $kT_{\rm e}\gtrsim$ 40 keV was estimated. The spectral plots corresponding to the best-fit models 1a and 1b are shown in Figure \ref{fig:relxill}.}\\

	\begin{table}
		\centering
		\caption{Best-fit spectral parameters for the \textit{NuSTAR} FPMA/B (Obs 1) of the source J17062 in energy 3-30 keV using Model 1a: \texttt{const*TBabs*(relxill)} and Model 1b: \texttt{const*TBabs*(relxillCP)}. Hydrogen column density ($N_{\rm H}$) for the component \texttt{TBabs} was fixed at $0.1\times10^{22}$ cm$^{-2}$ for the two models. All uncertainties are reported at a 90\% confidence level, computed using the MCMC chain fits within \textsc{XSPEC}.}
		\renewcommand{\arraystretch}{1.2}
		\begin{tabular}{ccc}
			\hline
			Component & Parameter (unit) & Value \\
			\hline
			\multicolumn{3}{c}{\small{Model 1a: \texttt{const*TBabs*(relxill)}}} \\
			\hline
			\small\textsc{constant} & \small{FPMB (wrt FPMA)} & \small{0.99$\pm$0.01} \\
			\textsc{relxill} & $i (^\circ)$ & 41$_{-4}^{+1}$ \\
			& $R_{\rm in}$ ($R_{\rm ISCO}$) & $\le3.6$\\
			& $\Gamma$ & 2.38$_{-0.03}^{+0.01}$ \\
			& log $\xi$ & 3.83$_{-0.06}^{+0.19}$ \\
			& $A_{\rm Fe}$ & 1.87$_{-0.49}^{+0.31}$\\
			& $E_{\rm cut}$ (keV) & $\ge$930 \\
			& \parbox{4cm}{Reflection fraction} & 1.11$_{-0.22}^{+0.33}$ \\
			& Norm (10$^{-3}$)& 3.02$\pm$0.04 \\
			\hline
			& $\chi^2/dof$ & 516/536 \\
			\hline
			\multicolumn{3}{c}{\small{Model 1b: \texttt{const*TBabs*(relxillCP)}}} \\
			\hline
			\small\textsc{constant} & \small{FPMB (wrt FPMA)} & \small{0.99$_{-0.02}^{+0.03}$} \\
			\textsc{relxillCP} & $i (^\circ)$ & 21$_{-11}^{+14}$ \\
			& $R_{\rm in}$ ($R_{\rm ISCO}$) & $\le2.6$\\
			& $\Gamma$ & 2.81$_{-0.10}^{+0.06}$ \\
			& log $\xi$ & 3.33$_{-0.01}^{+1.30}$ \\
			& log $N$ & 19.08$_{-0.87}^{+0.09}$ \\ 
			& $A_{\rm Fe}$ & 5.02$_{-2.36}^{+4.34}$ \\
			& $kT_{\rm e}$ (keV) & $\gtrsim$40 \\
			& \parbox{4cm}{Reflection fraction} & 0.14$_{-0.12}^{+0.66}$ \\
			& Norm (10$^{-3}$)& 3.10$_{-0.11}^{+0.22}$ \\
			\hline
			& $\chi^2/dof$ & 502/535 \\
			
			\hline
			
			\small\textit{Note:} & \multicolumn{2}{l}{\small{$dof$ refers to the degrees of freedom ascertained by the components for the best-fit model.}} \\
			\renewcommand{\arraystretch}{1.0}
		\end{tabular}
		\label{tab:pers_nustar}
	\end{table}
	
	
	We found that both flavors of reflection models described the 3-30 keV energy spectrum statistically equally well. Moreover, in both reflection models, we let the model return the primary continuum and the expected reflected spectrum from the disc. The resulting reflection spectrum depends on the nature of the illuminating (primary) continuum that provides the hard X-rays. The model \texttt{relxill} uses a power-law input spectrum, whereas \texttt{relxillCP} assumes that Comptonized disc emission irradiates the disc. We further note that the model \texttt{relxillCP} has a hard-coded seed photon temperature of 0.01 keV originating from the disc. Therefore, this self-consistent reflection model calculates the continuum spectrum assuming a seed temperature of $k{\rm {T_{seed}}}$ = 0.01 keV. Our continuum fitting with the model \texttt{nthcomp} revealed a disc temperature of 0.75 keV. Moreover, the \texttt{thcomp} model, an updated version of \texttt{nthcomp}, yielded even a higher disc temperature, around 1.41 keV. So, the hidden parameters of the \texttt{relxillCP} model may affect other continuum parameters and reflection parameters in unknown ways. This reflection model is usually preferred for spectral states where the seed-photon temperatures are low \citep{Ludlam2019, mondal2020disc}. This model may not be appropriate for the current spectral state.\\
	
\begin{figure}
		\centering
		\includegraphics[width=0.68\columnwidth, height=8.75cm]{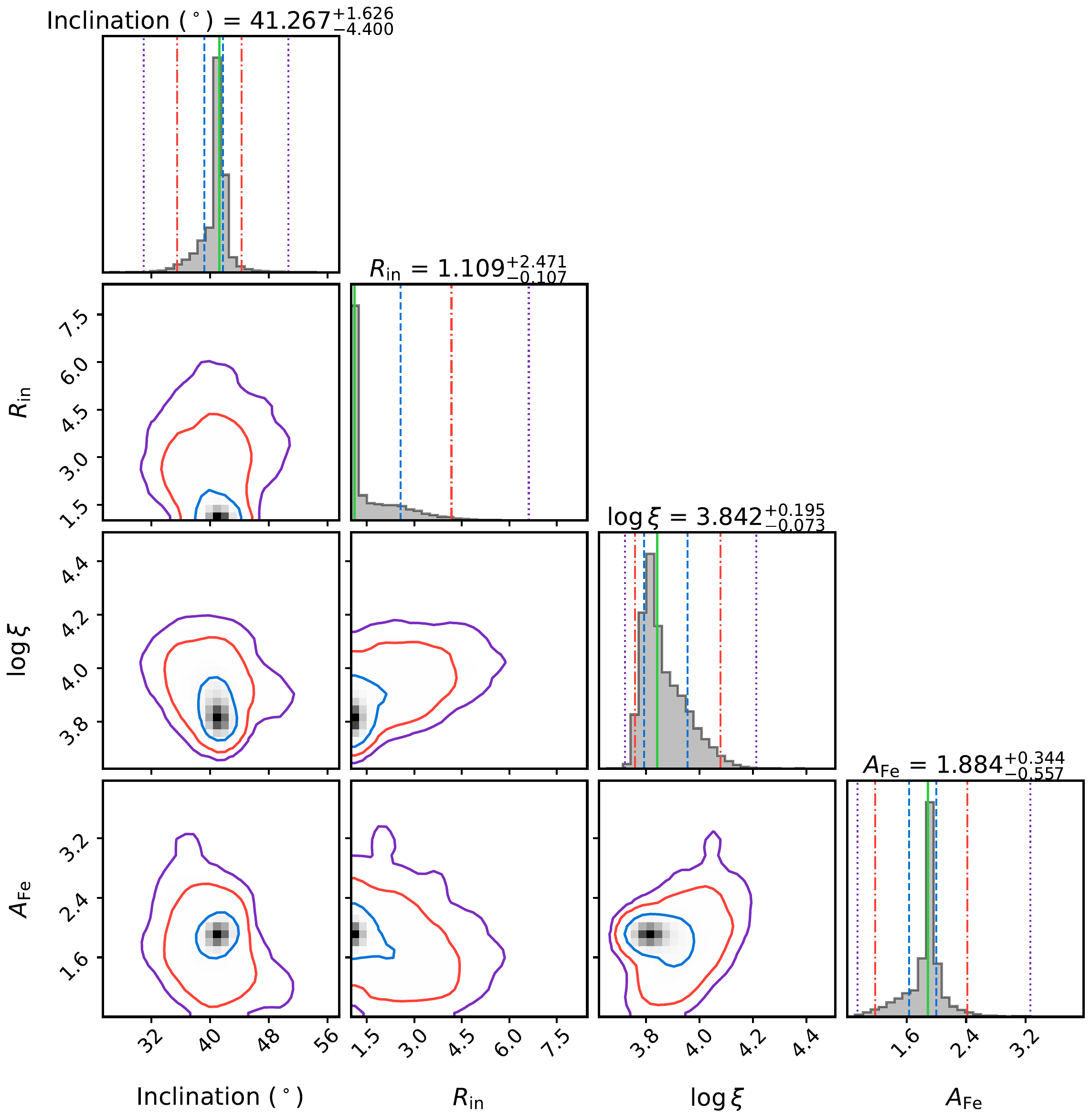}
		\caption{Corner plot using MCMC chain within \textsc{XSPEC} for Model 1a \texttt{const*TBabs*relxill} corresponding to the \textit{NuSTAR} observation (Obs 1). In the two-dimensional contour panels, the black marker denotes the best-fit parameters, while the blue, red, and violet colored contours represent the 1$\sigma$, 2$\sigma$, and 3$\sigma$ confidence levels, respectively. In the one-dimensional histogram panels, the solid green line indicates the best-fit value, and the blue dashed, red dot-dashed, and violet dotted lines signify the 1$\sigma$, 2$\sigma$, and 3$\sigma$ parameter uncertainties, respectively.}
		\label{relxill_mcmc}
	\end{figure}

	To further assess the robustness of the \texttt{relxill} model parameters, we performed Markov Chain Monte Carlo (MCMC) analysis, considering the uncertainties of the relevant reflection parameters \citep{foreman2013emcee}. The corner plot was generated using the \texttt{renorm} command with the Goodman-Weare chain type. The desired chain FITS files for both \texttt{relxill} model was generated using \textsc{xspec}. For Model 1a, we generated a chain of length 100,000 using 100 walkers with 8 parallel walkers. The resulting corner plot is shown in Fig. \ref{relxill_mcmc}. The corner plot corresponding to the uncertainties of the parameters of interest for the \texttt{relxillCP} model is not central to this analysis. Therefore, it is provided in \ref{app:mcmc} (Fig. \ref{fig:mcmc}).
	
	\subsection{\textit{NICER}}
	\subsubsection{Persistent emission}
	For Obs 3, we extracted the energy spectra for the persistent and burst emission (B1, B2, and B3). For persistent emission, we considered a total exposure of 9120 s, consisting of the interval from the beginning of the observation to 29,200 s and then from 55,835 s to 62,670 s. We generated associated background spectra using the \textsc{Scorpeon} background model. We found that the source is detected well above the background level up to 6.0 keV for the persistent emission. Hence, we performed the spectral fit over the 0.7-6.0 keV energy band. A similar background dominance was observed in the earlier \textit{NICER} observations performed in 2020 \citep{Bult_2021}. We initially modeled the persistent \textit{NICER} spectrum in the 0.7-6.0 keV range using the absorbed disk blackbody (\texttt{diskbb}) plus \texttt{power-law} model, \texttt{TBabs*(diskbb+powerlaw)}. The model estimated multi-color disk temperature $kT_{\rm in}\sim 0.36$ keV and photon index $\Gamma\sim 1$ with a fit statistics of $\chi^2/dof=173/82\sim 2.11$. To obtain a physically appropriate description, we modeled the spectrum without the \texttt{power-law} component, and instead convolved the disk blackbody component with the Comptonization model \texttt{thcomp}, \texttt{TBabs*(thcomp*diskbb)}. This model description fits the continuum emission well ($\chi^2/dof=171/81$), but we observed two prominent line emissions around 1 keV and around 2.3 keV. The 1 keV line was also present in earlier \textit{NICER} observations \citep{Bult_2021}. We added two \texttt{Gaussian} components to model the 1 keV and 2.3 keV emission lines. Adding two \texttt{Gaussian} line components to describe this feature improved the fit statistic by $\Delta \chi^2 = -98$, with 6 additional $dof$. Finally, this model \texttt{TBabs*(thcomp*diskbb+Gauss1+Gauss2)} combination described the overall emission well, with $\chi^2/dof=73/75$. The best-fit persistent spectral parameters are listed in Table \ref{tab:pers_burst}. The spectral plot is shown in the left panel of Figure \ref{fig:B1,B2,B3}.

	\begin{table}
		\centering
		\caption{Best-fit spectral parameters for the \textit{NICER} observation (Obs 3) of the source J17062 for the persistent time segment considered from the beginning of observation to 29,200s and then from 55,835s to 62,670s, culminating to a total of 9.12ks exposure, and the burst emission corresponding to the three bursts B1, B2 and B3. We used Model 2 and 3 for spectral fit of persistent and burst emission, respectively, where - Model 2: \texttt{TBabs*(thcomp*diskbb+Gauss1+Gauss2)} and Model 3: \texttt{TBabs*(thcomp*bbodyrad+Gauss2)}. The persistent and burst spectra are fitted over the energy band 0.7-6.0 keV and 0.7-5.0 keV, respectively. The unabsorbed flux ($F_{\rm unabs}$) for each spectra are reported corresponding to the 0.2-10.0 keV energy range.}
		\renewcommand{\arraystretch}{1.1}
		\begin{tabular}{cccccl}
			\hline
			Component & Parameter(unit) & Persistent(9120s) & B1(351s) & B2(350s) & B3(70s) \\
			\hline
			&&&&&\\
			\textsc{Tbabs} & $N_H (10^{22})$ $cm^{-2}$ & $0.11\pm0.02$ & 0.16$^{+0.07}_{-0.03}$ & 0.44$\pm$0.02 & 0.28$^{+0.13}_{-0.09}$\\
			\textsc{diskbb} & $kT_{\rm in}$ (keV) & $0.36\pm0.02$ &  &  &  \\
			& Norm & $39_{-5}^{+7}$ &  &  &  \\
			\textsc{bbodyrad} & $kT$ (keV) & & $0.26\pm0.02$ & $0.20\pm0.01$ & $0.26\pm0.03$ \\
			& Norm & & $3771_{-987}^{+1211}$ & $61234_{-2422}^{+28776}$ & $4176_{-2155}^{+7007}$ \\
			\textsc{thComp} & $\Gamma_{\tau}$ & $\le1.01$ & $\le1.01$ & $\le1.01$ & $\le1.04$ \\
			& $kT_e$ (keV) & $3.16_{-0.49}^{+0.51}$ & 2.19$_{-0.15}^{+0.12}$ & 2.13$_{-0.02}^{+0.15}$ & $2.55_{-0.59}^{+0.46}$\\
			& \parbox{1.2cm}{Covering \\ fraction} & $0.13\pm0.01$ & $0.60_{-0.09}^{+0.03}$ & $0.37_{-0.08}^{+0.02}$ & $0.47_{-0.05}^{+0.08}$ \\
			\textsc{Gaussian1} & \parbox{2.5cm}{$E_{\rm centroid}$ (keV)} & $0.99_{-0.09}^{+0.05}$ &  &  &  \\
			& $\sigma$ (keV) & $0.19\pm0.04$ &  &  &  \\
			& Norm ($10^{-4}$) & $8.60_{-3.0}^{+7.7}$ &  &  &  \\
			\textsc{Gaussian2} & \parbox{2.5cm}{$E_{\rm cenroid}$ (keV)} & $2.50\pm0.02$ & $2.34\pm0.02$ & $2.42\pm0.03$ & 2.29$\pm$0.05 \\
			& $\sigma$ (keV) & $0.18\pm0.04$ & $0.05\pm0.02$ & $0.09\pm0.02$ & $0.08\pm0.05$ \\
			& Norm ($10^{-5}$) & $9.61_{-2.46}^{+3.03}$ & $40.8_{-9.10}^{+10.60}$ & $1.60\pm0.40$ & $57.30_{-26.30}^{+28.1}$ \\
			\textsc{cflux} & $F_{\rm unabs}^\dagger$ & 8.19 & 72.60  & 124.78 & 38.36 \\
			\hline
			& $\chi^2$ /$dof$ & 73/75 & 83/55 & 137/60 & 63/46 \\
			\hline
			\textit{Note:} & \multicolumn{5}{l}{$dof$ refers to the degrees of freedom ascertained by the components for the best-fit model.} \\
			& \multicolumn{5}{l}{$^\dagger$ denotes the unit of flux $10^{-11}$ erg cm $^{-2}$ s $^{-1}$.}
			\renewcommand{\arraystretch}{1.0}
		\end{tabular}
		\label{tab:pers_burst}
	\end{table}
	
	\subsubsection{Burst spectral analysis}
	
	We extracted the spectra from all the type-II-like burst emission separately from Obs 3. Individual spectra were generated for each X-ray burst as listed in Table \ref{tab:Burst_times}, and analyzed the time-averaged spectra over the energy range of $0.7-5.0$ keV. Initially, we used the absorbed blackbody model, \texttt{TBabs*bbodyrad}, to fit the burst spectra. However, it failed to model the observed spectra, leading to higher reduced chi-squared values, irrespective of whether $N_{\rm H}$ was frozen or free during spectral fitting. We then tried an alternative model for the type-II X-ray burst: an absorbed Comptonization model, previously used by \cite{typeII_galloway} and \cite{type2_2026}. We applied \texttt{TBabs*thcomp*bbodyrad} to fit the burst spectra, considering that the seed photons from the neutron star surface/boundary layer are up-scattered in a Comptonizing corona. Burst spectra are well described by the absorbed thermal Comptonized model \texttt{TBabs*thcomp*bbodyrad}. We observed a residual around 2.3 keV, which we modeled with a \texttt{Gaussian} component. This feature was also present in the persistent emission. However, we have not detected the 1 keV emission line in the burst spectra. The reason for the non-detection of the 1 keV line is not clear. It may be due to limited statistics during bursts or to a physical process that weakens the line. All best-fit parameters for the burst spectral analysis within the energy range 0.7--5.0 keV, and burst fluxes in the energy range 0.2--10.0 keV are listed in Table \ref{tab:pers_burst}.
	
	
	However, a notable result from the fit of the persistent and time-averaged burst emission in the \textit{NICER} observation was the seed photon temperature during the persistent emission, characterized by a multi-temperature blackbody (\texttt{diskbb}) of $\sim$0.36 keV, which is slightly higher than the seed photon temperature of $\sim$0.26 keV, characterized by a single-temperature blackbody (\texttt{bbodyrad}), during the burst emission, usually interpreted as the emission from close to the neutron star surface or boundary layer. This discrepancy suggests that the seed photon during the persistent phase might be from the hotter part of the accretion flow/boundary layer, fed by mass accretion. At the same time, the Comptonized emission during the burst may be associated with the comparatively cooler part of the accretion flow. \\
	
	\begin{figure*}
		\centering
		\includegraphics[width=0.325\columnwidth,height=4.5cm]{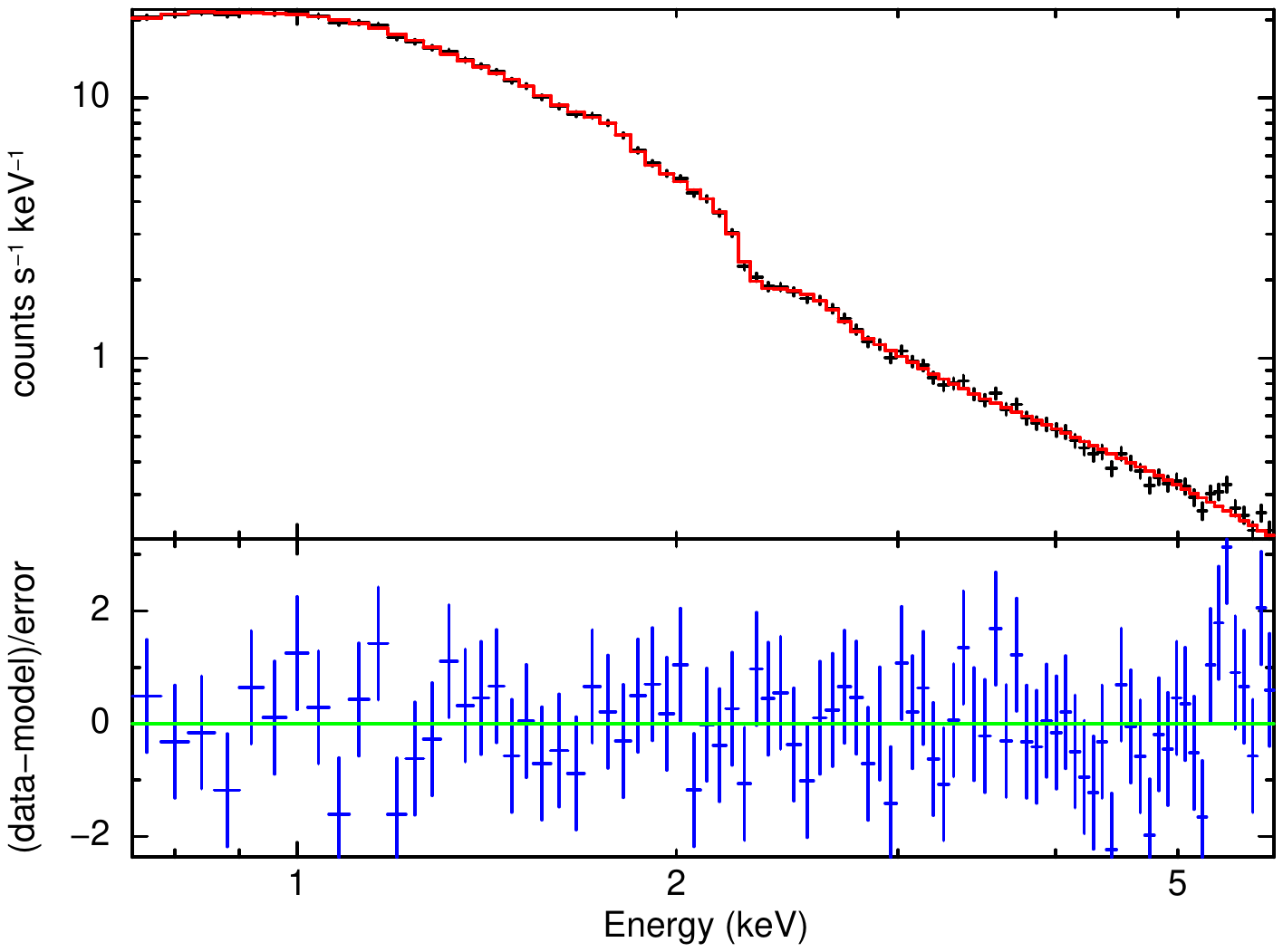}
		\includegraphics[width=0.325\columnwidth,height=4.5cm]{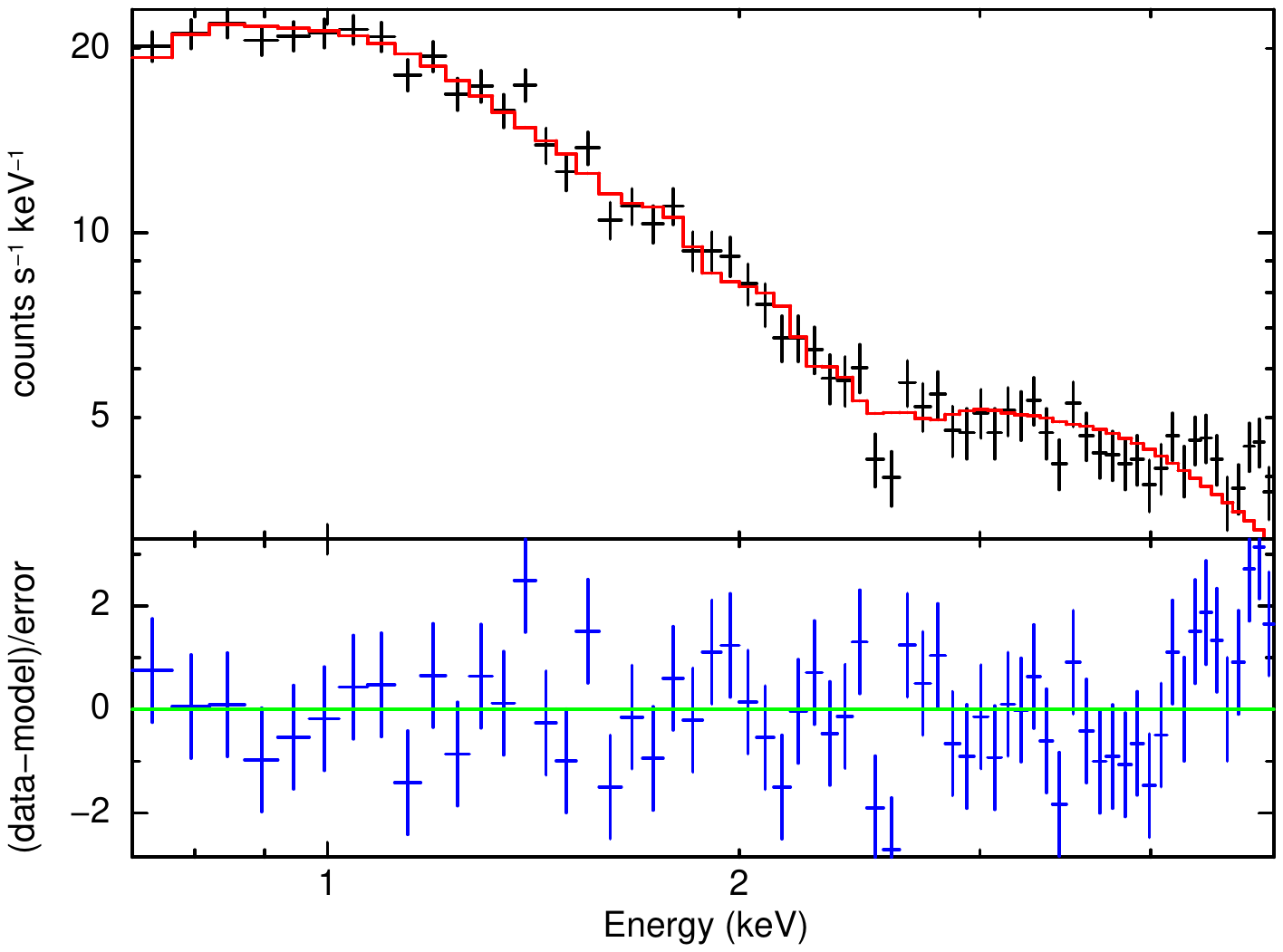}
		\includegraphics[width=0.325\columnwidth,height=4.5cm]{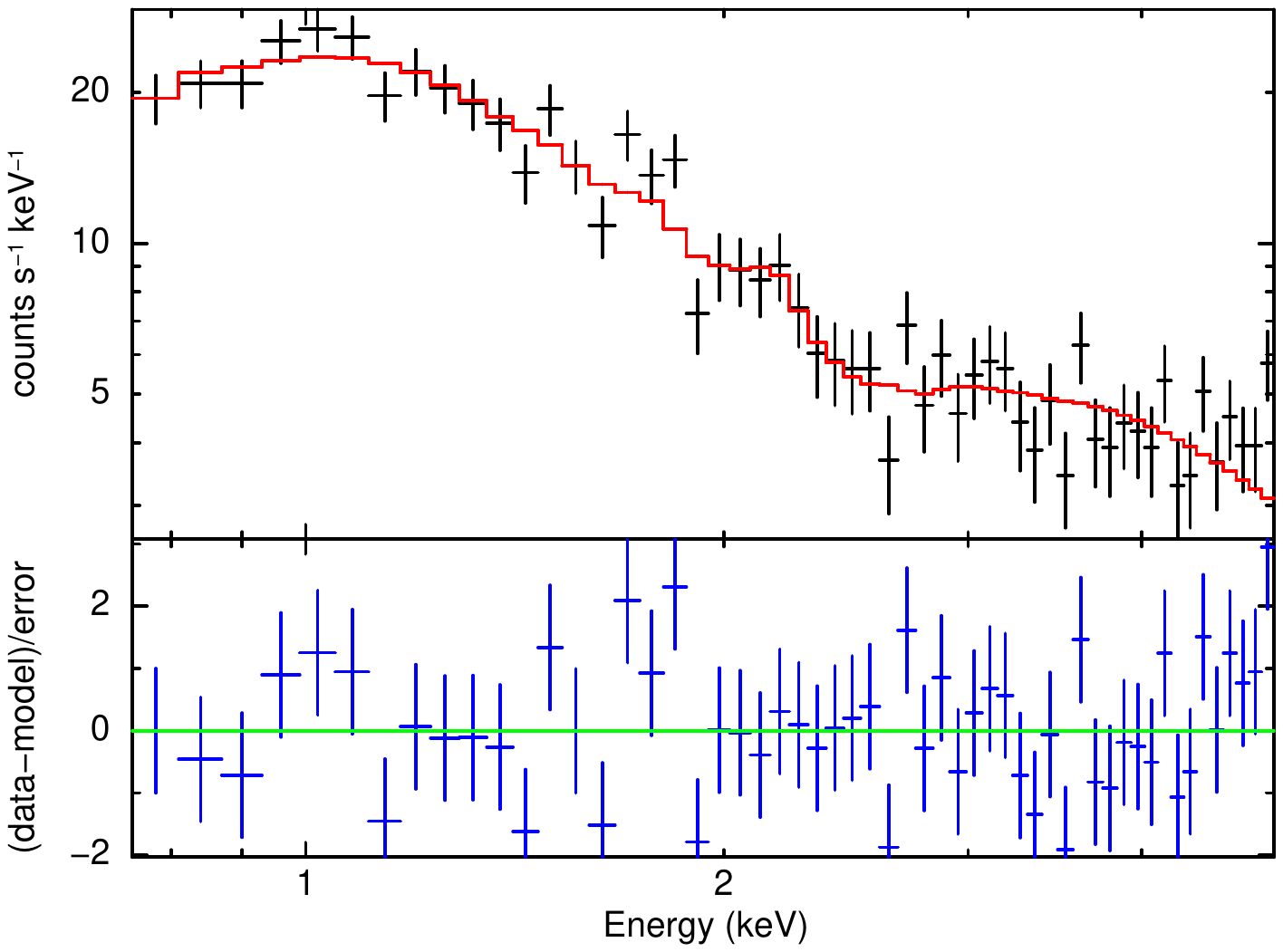}
		\caption{\textit{Left panel}: Unfolded spectra for the persistent emission for \textit{NICER} observation (Obs 3) for the best-fit values of the Model 2 -- \texttt{TBabs*(thcomp*diskbb+Gauss1+Gauss2)} for in the energy range 0.7-6.0 keV. \textit{Middle and right panel:} Unfolded spectra corresponding to the burst emission B1(\textit{middle panel}) and B3(\textit{right panel}) within the energy range 0.7-5.0 keV for the best-fit values corresponding to the Model 3 -- \texttt{TBabs*(thcomp*bbodyrad+Gauss2} for Obs 3 of the source J17062. The lower plots within each plot show the residuals with 1$\sigma$ error bar.}
		\label{fig:B1,B2,B3}
	\end{figure*}

	\subsection{Time-resolved analysis of the burst spectra}
	
\begin{figure*}
		\centering
		\includegraphics[width=0.6\columnwidth, height=12cm]{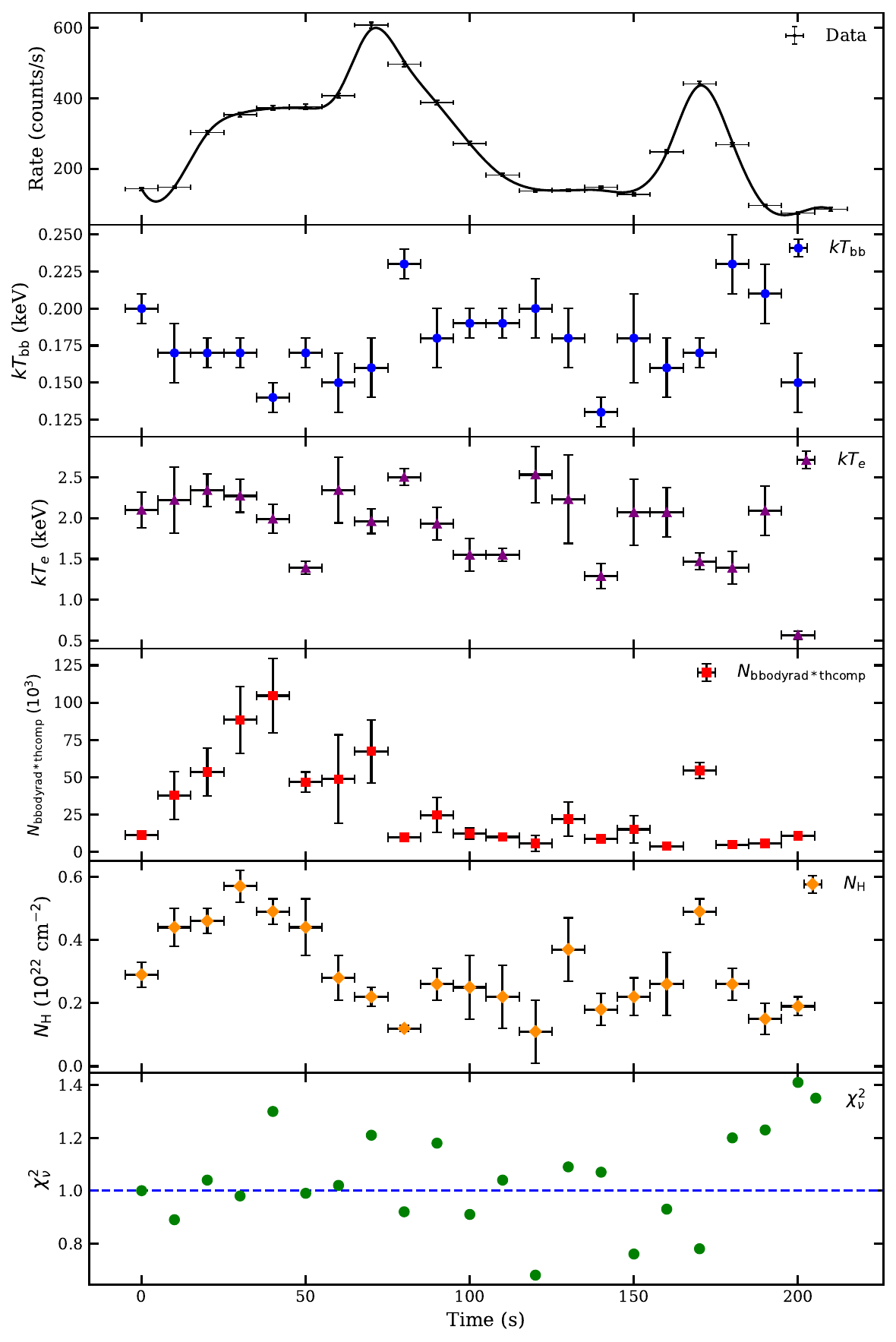}
		\caption{Light curve of B2 is shown in the first panel from MJD 59694.48. From the second panel onward, variation of the best-fit parameters of blackbody temperature ($kT_{bb}$), electron temperature ($kT_{\rm e}$), normalization of the Comptonized emission ($N_{\rm {bbodyrad * thcomp}}$), hydrogen column density ($N_{\rm H}$) and reduced $\chi^2$ corresponding to the Model 3 \texttt{TBabs*(thcomp*bbodyrad+Gauss2)} for the 10s segments  of B2 in the energy range 0.7-5.0 keV of Obs 3.}
		\label{fig:kT_time}
	\end{figure*}

	We performed the time-resolved spectral analysis of the most intense and energetic bursts (B2, peak count $\sim$850 counts/s) observed in the \textit{NICER} Obs 3 listed in Table \ref{tab:obs_all}. We intended to track the evolution of the Comptonized emission from close to the neutron star surface or boundary layer during the burst. Because the light curve of the burst shows a slow rise and fall after rapid recurring bursts (see Figure \ref{fig:bursts}), we excluded the first 140 seconds from our analysis. We divided the intermediate time interval into 21 segments, each 10 seconds long. We extracted spectra for each interval and attempted to fit them with the \texttt{TBabs*(thcomp*bbodyrad+Gauss2)} model, as applied to the time-averaged burst spectrum earlier. During the fitting the neutral hydrogen column density parameter of \texttt{TBabs} component was kept free. For individual sections, we also applied a \texttt{Gaussian} model. Initially, we let the model parameters of the \texttt{Gaussian} model be free during the fitting. However, we found that the parameter values are consistent with the persistent values and with those obtained from spectral analysis of the time-averaged burst analysis. The same methodology was applied to the \texttt{thcomp} model parameters. We let the power-law photon index ($\Gamma_{\tau}$), the electron temperature ($kT_{\rm e}$), and the covering fraction ($f_{\rm cov}$) free during the fitting. We found that the power-law photon index ($\Gamma_{\tau}$) is pegged at its lower value $\sim$1 and that the covering fraction ($f_{\rm cov}$) changes little, remaining within $\sim$0.3-0.6. Hence, we fixed $\Gamma_{\tau}$ and $f_{\rm cov}$, along with other parameters, to explore variations in the temperature of the seed photons ($kT_{\rm bb}$) and the electron temperature of the hot plasma ($kT_{\rm e}$), as well as the normalization of the Comptonized emission during the burst. The spectral fitting resulted in the seed photon temperature $kT_{\rm bb}\sim0.13 - 0.25$ keV and the electron temperature of the hot plasma $kT_{\rm e}\sim 1-3$ keV. The resulting parameter variations are shown in the Figure \ref{fig:kT_time}. We also note that the parameter variations are very similar to those measured after keeping the other Comptonized emission-related parameters free during the fitting.\\
	
	
	\section{Discussion}
	\label{sec:disc}
	
	We performed a spectral analysis of two observations of the accreting millisecond pulsar J17062 obtained with \textit{NICER} and \textit{NuSTAR} on 2022 April 25 and September 19, respectively. The unabsorbed persistent fluxes during the \textit{NuSTAR} (Obs 1) and \textit{NICER} (Obs 3) observations, within the energy range 3--79 keV and 0.2-10 keV, are 7.62$\times$10$^{-11}$ and 8.19$\times$10$^{-11}$ ergs cm$^{-2}$ s$^{-1}$, respectively. The persistent \textit{NuSTAR} flux is consistent with the 3-50 keV flux obtained from the earlier \textit{NuSTAR} observation studied by \cite{keek2017x}. The above-mentioned flux values correspond to the luminosities of $\sim$0.13\% and $\sim$0.14\% of $L_{\rm {Edd}}$, respectively (where the Eddington luminosity is given as $L_{\rm Edd}=3.8\times10^{38}$ ergs s$^{-1}$, considering the neutron star of mass M = 1.4M$_{\odot}$ and radius R=10 km). The estimated values of the fluxes, luminosities, and mass accretion rates (listed in Table \ref{tab:mass_acc}) during Obs 1 and 3 indicate that the source remained in the so-called hard spectral state during these observations. The unabsorbed bolometric flux in the 0.1-100 keV range for the \textit{NuSTAR} Obs 1 and \textit{NICER} Obs 3 are obtained as $\sim1.1\times10^{-10}$ and $\sim1.7\times10^{-10}$ ergs cm$^{-2}$ s$^{-1}$, respectively. We, further, estimated the corresponding bolometric luminosities as $\sim0.7\times10^{36}$ and $\sim1.1\times10^{36}$ erg s$^{-1}$, respectively. The continuum emission of the \textit{NuSTAR} spectrum in the energy band $3-30$ keV is entirely dominated by a power-law component or by Comptonized emission of disk photons of temperature $\sim 0.75$ keV by a plasma with a high electron temperature ($kT_{\rm e} \gtrsim 100$ keV).\\
	
	\begin{table*}
		\caption{Luminosity and mass accretion rate for \textit{NuSTAR} Obs 1 and \textit{NICER} Obs 3 were estimated corresponding to the unabsorbed fluxes in the 3--79 keV and 0.2--10 keV energy range, respectively, obtained using the \texttt{cflux} component in \textsc{XSPEC}. The values were calculated assuming a distance of $d\sim$7.3 kpc \citep{Keek2017}. The radius and mass of the source are assumed to be the standard values of 10 km and 1.4$M_{\odot}$.}
		\begin{tabular}{c c c c}
			\hline
			Observation Year and Month & \parbox{2.55cm}{Flux$^\%$ ($F$)\\ \small{(ergs cm$^{-2}$ s$^{-1}$)}} & \parbox{2.5 cm}{Luminosity$^{*}$ \\ \centering $ L = 4 \pi d^2 F$ \\ (ergs s$^{-1}$)} & \parbox{4cm}{Mass accretion rate \\ \centering $\dot{M} = \frac{LR}{GM}$ \\ \centering(g s$^{-1}$)} \\
			\hline
			2022 Sep (30801032002) & 7.62$\times$10$^{-11}$  & 4.85$\times$10$^{35}$  & 2.60$\times10^{15}$ \\
			2022 Apr (5531010105)$^*$ & 8.19$\times$10$^{-11}$ & 5.22$\times$10$^{35}$ & 2.80$\times10^{15}$ \\
			2022 Apr (5531010105 B1)$^\%$ & 5.78$\times$10$^{-10}$ & 3.68$\times$10$^{36}$ & 1.97 $\times 10^{16}$ \\
			2022 Apr (5531010105 B2)$^\%$ & 1.02$\times$10$^{-9}$ & 6.50$\times$10$^{36}$ & 3.48$\times10^{16}$ \\
			2022 Apr (5531010105 B3)$^\%$ & 2.86$\times$10$^{-10}$ & 1.82$\times$10$^{36}$ & 9.76 $\times 10^{15}$ \\
			\hline
			\multicolumn{4}{l}{\footnotesize$^*$ \textit{indicates value corresponding to persistent emission; $^\%$ indicates values corresponding to the peak burst emission}.}
		\end{tabular}
		\label{tab:mass_acc}
	\end{table*}

	In addition to continuum emission, prominent reflection features were detected, including a broad Fe K emission line around $5-8$ keV and a Compton hump around $15-30$ keV. The presence of such features strongly suggests a reflection of hard coronal photons from the accretion disk. We used the self-consistent relativistic reflection model \texttt{relxill} to investigate the parameters associated with the disk. The inner radius ($R_{\rm in}$) and inclination angle ($i$) of the accretion disk are primary to ascertain the truncation of the disk. Table \ref{tab:pers_nustar} shows all the best-fit values at the 90\% confidence level obtained by \texttt{relxill} model. From the spectral analysis for the \textit{NuSTAR} observation (Obs 1) using \texttt{relxill} (Model 1a), the angle of inclination ($i$) is ascertained as $\sim41^{\circ}$, which is consistent with \cite{Bult_2021}. The maximum inner radius of the disk from Model 1a is estimated to be 3.60 $R_{\rm ISCO}$, where  $R_{\rm ISCO}$ is the radius of the innermost stable circular orbit. Following \citealt{1998ApJ...509..793M} we have, $R_{\rm ISCO} = \frac{6GM}{c^2} [1 - j (\frac{2}{3})^{1.5}]$, where $M$ is the mass of the neutron star and $j$ is the dimensionless angular momentum $j\simeq cJ/GM^2$ with $J$ being the angular momentum of the star. For J17062, assuming the spin parameter of $j=0.1$, we obtain $R_{\rm ISCO} = 5.64 GM/c^2 \sim$ 11.69 $\frac{M}{M_{\odot}}$ km. Thus, the maximum values of inner disk radius is 20.30 $R_{\rm g}$, where $R_{\rm g}=\frac{GM}{c^2}\simeq2.07$ km (assuming $M=1.4 M_{\odot}$), for Model 1a. The inward extension of the accretion disk implies accreting matter reaching the neutron surface \citep{2006MNRAS.369.2036S}. Our reflection fit with \texttt{relxill} returned a high value of $E_{\rm cut}$ above 930 keV, consistent with \cite{keek2017x}.\\
	
	From the modeling of the reflection spectrum, \cite{Degenaar2017} suggested that the accretion disk is truncated far from the neutron star surface at $\gtrsim 100 R_{\rm g}$ when the source is accreting at $0.1$ percent of the Eddington limit. The result was consistent with the different choice of the inclination angle (as it was not well constrained by the data). However, they also noted that $R_{\rm in}$ was consistent with a location at the $ISCO$ at 3$\sigma$ confidence for $i=65^\circ$. From the reflection studies, \cite{Keek2017} suggested that the inner disk is truncated at $\sim100 R_{\rm g}$ before the burst, but may move closer to the star during the burst. Although \cite{vandenEijnden2018} detected a truncated disc at around $77 R_{\rm g}$, assuming a high inclination angle, they could not rule out the possibility of a disk extending to the neutron star at low inclination. Our spectral studies suggested a disk extending close to the neutron star surface at low inclination, as revealed by a couple of the self-consistent reflection models. Therefore, the results are consistent with the earlier predictions of different authors based on reflection studies. However, in the low accretion rates, it is difficult to definitively distinguish whether the NS’s magnetic field truncates the disk, or the formation of a hot inner flow results in a large inner disk radius. This indicates a substantial change in the accretion geometry of the system over time. Nevertheless, the greater uncertainties associated with the earlier results, along with the present inference of a disk extending close to the NS rather than exhibiting a large truncation from two reflection models, may indicate that the previously reported disk estimates were overestimated owing to systematic biases.\\ 
	
	The photon indices obtained from the \texttt{relxill} model $\Gamma=$2.38$_{-0.03}^{+0.01}$,  consistent with the value of $\Gamma\sim$2.25$\pm$0.01 reported from the 2014 October \textit{Chandra} observation \citep{Keek2017}. The log $\xi$ parameter, which signifies the logarithm of ionization of the accretion disk, is estimated at $\sim$3.83 for the reflection model \texttt{relxill}, indicating a high ionization in the accretion disc. \cite{Degenaar2017} estimated $\log \xi\sim3.24$ and $\sim3.35$ for \textit{NuSTAR}/\textit{Swift} observations across two angles of inclination $25^{\circ}$ and 45$^\circ$, respectively. The accretion disk density ($N$) is estimated at the highest value $\sim 10^{19}$ cm$^{-3}$. The iron abundance ($A_{\rm Fe}$) of the accretion disk was found to be 1.87$_{-0.49}^{+0.31}$ times the solar value, indicating a moderately high iron abundance in the disc. We note that our analysis precisely constrains the value of $A_{\rm Fe}$, whereas previous studies fixed it to unity \citep{Degenaar2017}.\\
	
	The persistent emission corresponding to the \textit{NICER} Obs 3 were analysed using the \texttt{TBabs*(thcomp*diskbb+Gauss1+Gauss2)} model in the energy range $0.7-6.0$ keV. The spectral analysis for the persistent emission estimated the inner disk temperature as $T_{\rm in}\sim$0.36 keV. We obtained an emission line around $\sim$2.2 keV, as reported earlier by \cite{vandenEijnden2018}, during the persistent spectral fit. However, the origin of this line is not clearly understood. Apart from this, for the first time, five type-II-like bursts were detected during Obs 3 (see lower panel of Fig. \ref{fig:all_lc}). To date, only two sources exhibit type-II bursts: one is a bursting pulsar (BP), GRO J1744-28 \citep{kouveliotou1996new}, and the other is a rapid burster (RB), MXB 1730-335 \citep{lewin1976discovery}. Considering our detection of type-II-like bursts from the accreting pulsar J17062, it may become the third object from which type-II bursts have been detected, contingent on confirmation of the classification. Previously, long and energetic type-I thermonuclear X-ray bursts have been detected by $Swift$/XRT and $MAXI$/GSC, which decay as straight power laws, described by a single linear function with slopes of $-0.32$ to $-1.15$ \citep{2013ApJ...767L..37D, keek2017x}. Upon confirmation of the reported type-II-like bursts, the source may become only the second known accreting pulsar to exhibit both types of bursts. To understand the origin and behavior of type II X-ray bursts, a series of models have been proposed, focusing on the sudden release of the accretion power in the disk. These include disk–magnetosphere interactions \citep{Jiang_2014}, trapped disks or the magnetic barrier \citep{dangelo_2012}, instabilities in the accretion flow, and thermal or viscous instabilities in the accretion disk \citep{1974ApJ...187L...1L, 1976MNRAS.175..613S}.\\

	We have analysed the time-averaged burst spectrum corresponding to the three bursts B1, B2, and B3 using the model \texttt{TBabs*(thcomp*bbodyrad+Gauss2)} in the energy range $0.7-5.0$ keV. From time-averaged spectrum, the blackbody temperature for B1, B2, and B3 are estimated as $\sim$0.26, $\sim$0.20, and $\sim$0.26 keV respectively. Time-resolved spectral analysis of B2 (Figure \ref{fig:kT_time}) revealed that the electron temperature ($kT_{\rm e}$) of the corona and the seed photon temperature ($kT_{\rm bb}$) showed no significant variability during the burst. The parameters $kT_{\rm e}$ and $kT_{\rm bb}$, in general, vary between $1-3$ keV and $0.13-0.25$ keV, respectively. The neutral hydrogen column density parameter ($N_{\rm H}$) changes little but did not exhibit any specific trend during the burst. For most segments, the reduced $\chi^2$ remained very close to unity or lower. The lack of significant variability in the temperatures signifies that the burst emission may not originate in the neutron surface and possibly involves inner accretion flow, an extended emission or reprocessing region, or a boundary layer. No clear monotonic cooling trend is detected within the analysed portion of B2 under the adopted Comptonization model. The corona temperature remained roughly $\sim$2 keV, consistent with other \textit{ASCA} \citep{mahasena2003new} and \textit{RXTE} \citep{typeII_galloway} observations. As our \textit{NuSTAR} observation suggests a Comptonizing coronal region with an electron temperature $\gtrsim 100$ keV, these results support the idea of an additional, temporary coronal region with a temperature around 2 keV. Although the exact mechanism by which such a region forms is not clear, it may be linked with the magnetic reconnection on the disk \citep{Chen_2021}.\\
	
	
	The inward motion of the accreting material is governed by the competition between the ram pressure of the accretion flow and the magnetic pressure of the neutron-star magnetosphere. The disk is truncated at the magnetospheric radius ($R_{\rm m}$), where the ram pressure of the inflowing matter becomes comparable to the magnetic pressure \citep{Sunyaev_1975}. Following \cite{ghosh1978accretion}, magnetospheric radius is given as $R_{\rm m} \simeq \xi \left( \frac{\mu^4}{2G M \dot{M}^2}\right)^{1/7}$ where $\xi$ is the ratio of magnetospheric to Alfv$\acute{e}$n radius ($\xi\sim0.5-1.0$), $\dot{M}$ is the mass accretion rate, and $\mu$ is the magnetic moment of the neutron star. Following \citealt{2009MNRAS.400..492I}, using the distance to source as D=7.3 kpc, the magnetic moment can be calculated as
	\begin{equation}
		\begin{split}
			\nonumber
			\mu = 3.5 \times 10^{23} x^{7/4} k_{\rm A} ^{-7/4} \left( \frac{M}{M_{\odot}}\right)^2 \left( \frac{f_{\rm ang}}{\eta} \right)^{1/2}  \left( \frac{F_{\rm b}}{10^{-9} {\rm erg} \, {\rm cm}^{-2} \, {\rm s}^{-1}}\right)^{1/2} \left( \frac{D}{7.3 {\rm kpc}}\right)
		\end{split}
	\end{equation}
	
	where $\eta$ is the accretion efficiency, $f_{\rm ang}$ is the anisotropy correction factor, $F_{\rm b}$ is the flux and $k_{\rm A}$ is the geometric coefficient. Considering \cite{shakura1973black}, $\eta$ is given as, $L=\eta \dot{M}c^2$, where $L$ is the luminosity. As $f_{\rm ang}=\frac{1}{{\rm cos} i}$, the corresponding values of $f_{\rm ang}$ for \texttt{relxill} and \texttt{relxillCP} models are obtained as $\sim$1.31 and $\sim$1.06, respectively. Following \citealt{Degenaar2017}, we have assumed $k_{\rm A}=1$, $\xi=1$, $\eta=0.1$ and the factor $x$ is inferred from $R_{\rm in}$ = $\frac{x GM}{c^2}$ as $x\lesssim$ 20.3 for the upper limit corresponding to the \texttt{relxill} model (Fig. \ref{relxill_mcmc}). Corresponding to the value of flux Obs 1 from Table \ref{tab:mass_acc}, the maximum value of magnetic moment corresponding to the maximum value of $R_{\rm in}$ as inferred from the \texttt{relxill} model is estimated as $\sim4.2\times10^{25}$ G cm$^3$. The corresponding magnetic field is given as $B=\frac{\mu}{R^3}$. As we consider a neutron star of radius R=10km, the maximum value of magnetic field corresponding to \texttt{relxill} model is estimated as $\sim4.2$ $\times$ 10$^7$ G which is consistent with the value reported by \cite{Keek2017}. We consider $\xi=1$, then, substituting for the corresponding value of magnetic moment and mass accretion rate, we get the maximum value of $R_{\rm m}$ as $R_{\rm m,1}\sim$27.6 km for the \texttt{relxill} model. In contrast, the co-rotation radius ($R_{\rm co}$) is defined as the radius at which the Keplerian orbital frequency equals the neutron-star spin frequency. The co-rotation radius of the source can be computed using $R_{\rm co}=(\frac{GM}{\Omega^2})^{1/3}$ where $\Omega=2\pi f$ is the angular velocity corresponding to the pulsation frequency ($f$). For $f$=163.65 Hz and M$\sim$1.4M$_{\odot}$, we get $R_{\rm co}\sim$56.13 km. The comparison between $R_{\rm m}$ and $R_{\rm co}$ shows that $R_{\rm m} < R_{\rm co}$, which suggests the system to be in the ongoing accretion regime \citep{ghosh1979accretiona}. Assuming a mass of 1.4$M_{\odot}$, the inner disk radius inferred from spectral modeling of \texttt{relxill} ($R_{\rm in} \lesssim$ 42 km) lies well inside the co-rotation radius. These results indicate that accretion onto the neutron-star surface can proceed.
	Although \cite{vandenEijnden2018} suggested J17062 was operating near a propeller-dominated regime based on 2015 observations, the smaller inner disk obtained from the 2022 \textit{NuSTAR} observation suggests that the source was in an active accretion regime during our observation.\\
	
	The \textit{NICER} and \textit{NuSTAR} observations were performed in two different epochs and differ in the flux values. Our \textit{NICER} analysis suggests that at least some of the light curve profiles are associated with candidate type-II-like bursts. To explain the nature of the type-II-like bursts, we considered the accretion conditions during the \textit{NICER} observations. The spin of the NS is not expected to vary considerably over the interval of five months between \textit{NICER} and \textit{NuSTAR} observations. Hence, we adopt the value of co-rotation radius as $R_{\rm co}\sim56$ km for Obs 3. Similarly, the magnetic moment of the NS is expected to remain unchanged over the five-month interval. As the magnetospheric radius ($R_{\rm m}$) can vary with the accretion rate according to $R_{\rm m} \propto \mu^{4/7} \dot{M}^{-2/7}$. We can denote the parameters corresponding to the two epochs by the subscript showing the particular observation. We could not compute the inner disk radius during Obs 3 because the data were insufficient to incorporate reflection modeling.  However, the 3--79 keV \textit{NuSTAR} flux and the 0.2--10 keV \textit{NICER} flux do not represent the same fraction of the source luminosity and therefore cannot be directly compared to infer the ratio of mass--accretion rates. We therefore calculated the unabsorbed bolometric fluxes in the energy range 0.1-100 keV to estimate mass accretion rates for both observations (Obs 1 and Obs 3) and specifically to scale $R_{\rm m}$. Considering the bolometric fluxes, the mass accretion rates ($\dot{M}$) for the \textit{NuSTAR} Obs 1 and \textit{NICER} Obs 3 are obtained as $\sim3.8\times10^{15}$ and $\sim5.9\times10^{15}$ g s$^{-1}$, respectively. Incorporating the bolometric flux yielded a small change in the magnetic moment ($\mu$) for Obs 1. The magnetic moment ($\mu$) obtained for the bolometric flux for Obs 1 is $\sim5.1\times10^{25}$ G cm$^3$. Now, comparing with the obtained value for Obs 1, $R_{\rm m,1}\simeq 27.5$ km (considering bolometric flux), the magnetospheric radius during Obs 3 is obtained as $R_{\rm m,3}\sim24.3$ km. So, during Obs 3, the source remain in the accretion regime as $R_{\rm m,3} < R_{\rm co}$. We therefore used this magnetospheric radius to assess the accretion regime and do not equate it directly with an inner disk radius.\\
	
	Furthermore, the average recurrence time ($\Delta t_{\rm rec}$) for the three bursts (B1, B2 and B3) is obtained as $5571.35$ s (see Table \ref{tab:Burst_times}). The flux for the persistent emission during Obs 3 over the energy range $0.2-10$ keV is $F_{\rm pers}=8.19\times10^{-11}$ ergs cm$^{-2}$ s$^{-1}$. Considering the distance to the source as $d\sim$7.3 kpc, the persistent luminosity can be estimated as $L_{\rm pers}=4 \pi d^2 F_{\rm pers}=5.22\times10^{35}$ ergs s$^{-1}$, which corresponds to 0.001 $L_{\rm Edd}$. The flux for the persistent subtracted burst emission estimated over the energy range 0.2-10.0 keV for the five bursts are shown in the Table \ref{tab:spect_best_fit}. The gravitational energy released through accretion between bursts can be approximated as $E_{\rm acc} = L_{\rm pers} \Delta t_{\rm rec}$, where $\Delta t_{\rm rec}$ is the burst recurrence time. For $\Delta t_{\rm rec}=5571.35 s$, we obtain $E_{\rm acc}=2.91\times10^{39}$ ergs. For B1 and B2, the energy released during the burst ($E_{\rm burst}$) is comparable to the gravitational energy accumulated during the recurrence interval ($E_{\rm acc}$). The burst fluence ($f_{\rm b}$) represents the total energy per unit area emitted during the burst, and is calculated by integrating the burst flux, after subtracting the persistent flux, over the duration of the burst \citep{galloway2008thermonuclear}. During each burst, we estimate $f_{\rm b}$ by multiplying the time-averaged, persistent subtracted $0.2-10.0$ keV flux with the corresponding burst exposure as given in Table \ref{tab:Burst_times}. Moreover, we have considered the peak burst flux and the minimum flux corresponding to each burst to determine the spectral-flux uncertainties in the fluences. The peak burst fluence is computed using the integrated times of 83 s, 24 s, and 18 s for B1, B2, and B3, respectively. The minimum fluence of each burst is estimated by considering a segment of approximately 100 s from the onset of the burst. The ratio of the persistent fluence accumulated between bursts to the burst fluence is estimated by the parameter $\alpha$, which is key to distinguishing between type-I and type-II bursts. The nominal $\alpha$ value is calculated using the time-averaged burst fluence ($f_{\rm b}$) while the upper and lower limits of $\alpha$ correspond to the minimum and peak burst fluences, respectively. For the bursts B1, B2 and B3, we have obtained $\alpha$ within $\sim$0.9-19 based on time-averaged flux. As listed in Table \ref{tab:spect_best_fit}, the estimated $\alpha$ values for B1, B2, and B3 remain below 20, even after accounting for uncertainties in their fluences. We further note that, B1 would have $\alpha \simeq 1.9$, assuming a recurrence time comparable to those of the subsequent bursts. The consistent low values obtained for B1, B2, and B3 may be indicative of an accretion-powered origin of the bursts. Such low values of $\alpha$ are inconsistent with thermonuclear (type-I) bursts, for which the typical values lie within $40-100$. On the contrary, type-II bursts are essentially characterized by a low $\alpha$ value (below 20) that signifies the bursts being powered by episodic accretion \citep{type2_2026}. The overall result is suggestive of accretion-powered type-II-like bursts for the source IGR J17062.\\
	
	Moreover, type-II bursts have a long waiting time between bursts, which is needed for the accumulation of fuel. As demonstrated by \cite{type2_2026}, type-II bursts are often characterized by a positive correlation between burst fluence ($f_{\rm b}$) and the waiting time ($\Delta t$), reflecting the accumulation of accreting material in a reservoir prior to its sudden release. It is referred to as relaxation-oscillator behaviour and is considered a characteristic signature of type-II bursts. We consider the waiting time for each burst as the time interval between successive bursts identified from the full light curve of the \textit{NICER} Obs 3. By correlating the waiting times of B3, B4 and B5 with the respective burst fluences from Table \ref{tab:spect_best_fit}, we obtained the Pearson correlation coefficient (see \ref{app:pears}). As there are three bursts, we get $n=3$ and considering the waiting time between bursts and burst fluences for $i=1,2,3$ as variables $x_i$ and $y_i$, the Pearson correlation coefficient is obtained as $r=0.97$, which suggests a positive correlation between waiting time and burst fluence. The apparent positive correlation is a qualitative and exploratory result. However, the $p$-value derived from the Student's $t$-statistic is $0.15$ due to very small sample size. The available data show a tentative trend consistent with relaxation--oscillator behaviour, but the incomplete burst coverage and small sample prevent a detailed quantitative test.\\	
	
	\begin{table*}
		\centering
		\caption{Indicative parameters for the burst calculated on the basis of the spectral parameters obtained corresponding to the best-fit spectral Model~3: \texttt{TBabs*(thcomp*bbodyrad+Gauss2)} associated to the burst emission for the \textit{NICER} observation (Obs~3) of J17062 in the energy range 0.2--10.0~keV, where $E_{\rm burst}$ is the total energy released during the burst, $F_{\rm burst}$ is the flux corresponding to the full burst, $F_{\rm peak}$ is the peak burst flux, $f_{\rm b}$ is the burst fluence for the whole burst, $f_{\rm b, peak}$ is the peak burst fluence, $F_{\rm min}$ is the minimum flux during the initial rise of the burst, $\alpha$ is the ratio of integrated persistent emission to burst fluence, and $\Delta t_{\rm rec}$ is the average burst recurrence time.}
		\label{tab:spect_best_fit}
		\renewcommand{\arraystretch}{1.1}
		\begin{tabular*}{\textwidth}{@{\extracolsep{\fill}} l c c c c c c c c}
			\hline
			\hline
			\noalign{\smallskip}
			\shortstack[c]{Burst\\Serial\\no.} & 
			\shortstack[c]{$ F_{\rm burst}^\dagger$} & 
			\shortstack[c]{$f_{\rm b}^\$$} & 
			\shortstack[c]{$F_{\rm peak}^\%$} & 
			\shortstack[c]{$f_{\rm b, peak}^\$$} & 
			\shortstack[c]{$F_{\rm min}^{\dagger\#}$} & 
			\shortstack[c]{$E_{\rm burst}$\\$(4 \pi d^2f_{\rm b})$\\($10^{39}\text{ erg}$)} & 
			\shortstack[c]{$\alpha$\\$\left(\frac{F_{\rm pers} \Delta t_{\rm rec}}{f_{\rm b}}\right)$} & 
			\shortstack[c]{Range$^r$\\ of $\alpha$} \\
			\noalign{\smallskip}
			\hline
			\noalign{\smallskip}
			B1      & 7.26  & 25.4 & 5.87  & 48.7 & 2.82 & 1.62 & 1.9  & 1--18    \\
			B2      & 12.48 & 43.8 & 15.23 & 36.6 & 2.95 & 2.79 & 1.0  & 0.9--15  \\
			B3      & 3.84  & 2.7  & 1.38  & 2.5  & 3.55 & 0.17 & 16.7 & 10--19   \\
			B4$^*$  & 1.55  & 7.4  &  &  &  &  &  &    \\
			B5$^*$  & 1.36  & 3.9  &  &  &  &  &  &    \\
			\noalign{\smallskip}
			\hline
		\end{tabular*}
		\vspace{1ex}
		\begin{minipage}{\textwidth}
			\footnotesize
			\textit{Notes.} 
			$^\dagger$ Unabsorbed flux in units of $10^{-10}$ $\text{erg}\text{ cm}^{-2}\text{ s}^{-1}$; $^\%$ flux in units of $10^{-9}$ $\text{erg}\text{ cm}^{-2}\text{ s}^{-1}$. $^\$$ Fluence in units of $10^{-8}$ $\text{erg}\text{ cm}^{-2}$.	$^\#$ Minimum burst fluence is estimated using an integration timescale of 100~s. $^r$ Range of $\alpha$ is calculated using minimum and peak burst fluences. $^*$ Bursts B4 and B5 values are included for correlation analysis only. For B4 and B5, the tabulated fluences should be regarded as lower limits rather than measurements of the complete burst fluence, as the rise and peak phases were not completely observed for them.
		\end{minipage}
	\end{table*}
	
	During Obs 1, both the magnetospheric radius and the inner disk radius are smaller than the co-rotation radius, suggesting that the system remains in an accreting regime with ongoing disk-magnetosphere interaction. During Obs 3, we obtain a similar magnetospheric radius to that in Obs 1, assuming the magnetic moment remains constant between the two epochs. It indicates that the magnetospheric radius $R_{\rm m,3}$ moved well inside $R_{\rm co}$. In this scenario, as indicated by the inferred geometry, disk-magnetosphere interaction may be a possible interpretation of the observed type-II-like bursts from the source J17062. Observationally, this magnetospheric instability can produce bursts on a wide range of time scales \citep{dangelo_2012, Jiang_2014}. The observed characteristics of type-II bursts in our analysis may be attributed to episodic accretion instabilities driven by disc-magnetosphere interactions. Such a process could produce a short-lived increase in the accretion rate (see Table \ref{tab:mass_acc}), and may consequently give rise to bursts of X-ray emission. However, the exact mechanism by which the accretion material releases its potential energy in the innermost disk or near the NS surface remains unclear.  Consequently, the gravitational energy released through accretion between bursts, estimations of low $\alpha$ values, and the qualitatively apparent positive correlation between waiting time and burst fluence, suggest an accretion-driven origin of the bursts. However, further studies of burst properties and disk-magnetosphere interactions are required to establish whether the observed bursts are indeed type-II-like bursts and distinguish them conclusively from type-I bursts.
	
	\section{Data Availability}
	We used the archival data from the \texttt{NASA}'s \texttt{HEASARC} database.
	
	\section{Acknowledgments}
	We sincerely acknowledge the esteemed referee for his erudite comments, which led to the overall enhancement of the quality of the manuscript in all aspects. This research has made use of the \textit{NuSTAR} data analysis software \texttt{NuSTARDAS} jointly developed by the ASI Space Science Data Center (SSDC, Italy) and the California Institute of Technology (Caltech, USA). Additionally, \texttt{NICERDAS} was utilized for \textit{NICER} data reduction and analysis. For the analysis of archival data, we deeply acknowledge \texttt{HEASOFT} and \texttt{CALDB}. This research has made use of \textit{MAXI} data provided by \textit{RIKEN}, \textit{JAXA}, and the \textit{MAXI} team. We have used the \texttt{emcee} module available at \texttt{astropy} to create the corner plot. This research work was supported by the UGC non-NET fellowship grant of the Government of India. MB and SJ acknowledge the research programme and facilities provided by the Department of Physics, Visva-Bharati University. ASM and BR would like to thank Inter-University Centre for Astronomy and Astrophysics (IUCAA) for their facilities extended to them under their Visiting Associate Programme.
	
	\appendix
	\section{Corner plot}
	\label{app:mcmc}
\begin{figure}[!h]
		\centering
		\includegraphics[width=0.6\columnwidth, height=10 cm]{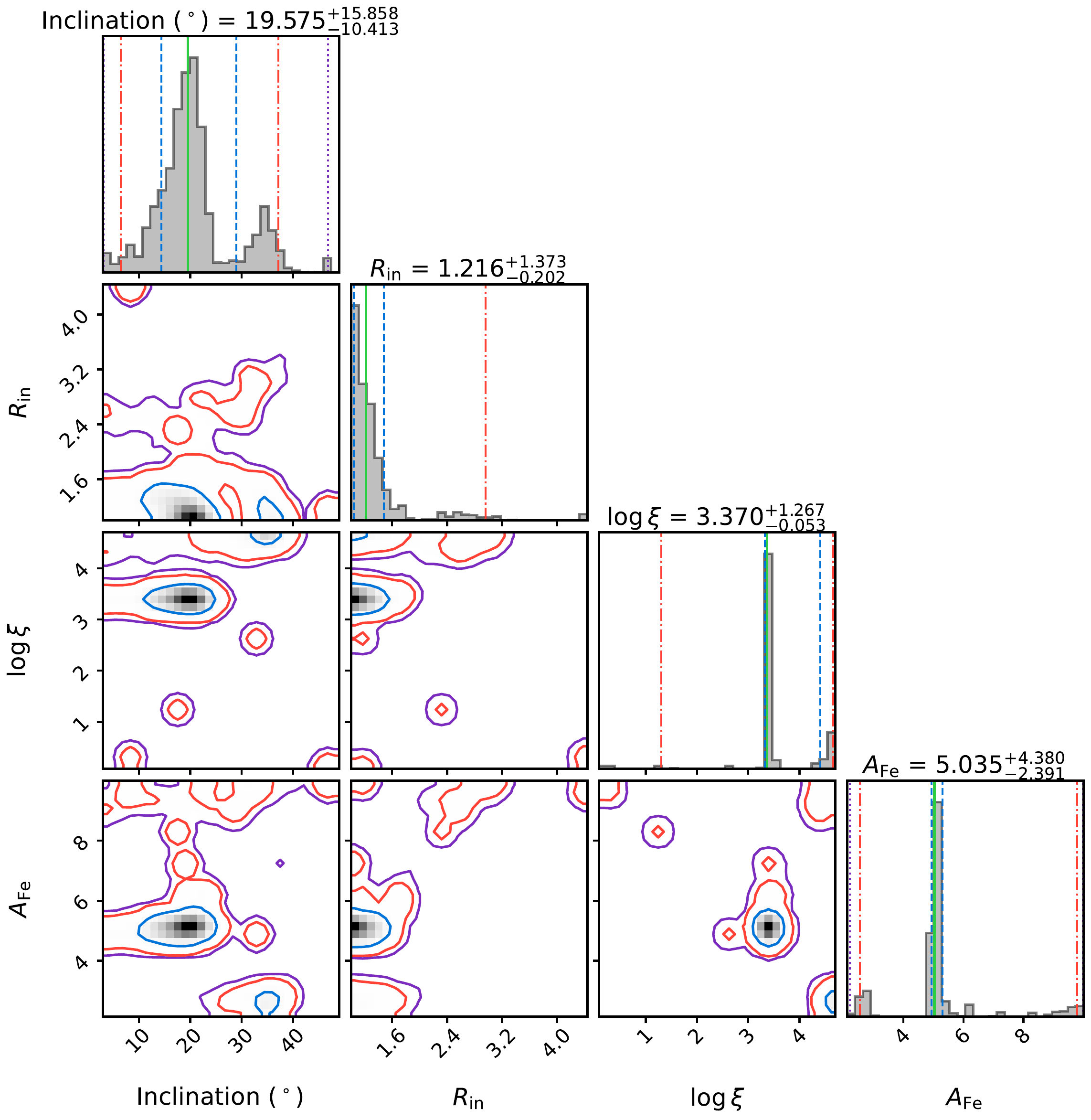}
		\caption{Corner plots using MCMC chain \textsc{.fits} file for Model 1b, \texttt{const*TBabs*relxillCP}, corresponding to the \textit{NuSTAR} observation (Obs 1). For each plot within the histogram panels, the middle solid green line corresponds to the best-fit value and the other three blue dashed, red dot-dashed, and violet dotted lines signify the 1$\sigma$, 2$\sigma$, and 3$\sigma$ parameter uncertainties, respectively. Similarly, for the two-dimensional contour panels, the central black region signify the best-fit value whereas the blue, red, and violet contours represent the 1$\sigma$, 2$\sigma$, and 3$\sigma$ confidence levels.}
		\label{fig:mcmc}
	\end{figure}

	The  Markov Chain Monte Carlo (MCMC) plot for the best fit model 1b was performed and shown in Figure \ref{fig:mcmc} for the crucial parameters \citep{foreman2013emcee}. The corner plot was created using the \texttt{renorm} command and the Goodman-Weare chain type. The desired chain FITS file for \texttt{relxillCp} model was created in \textsc{xspec}. A chain length of 100,000, along with 100 chain walkers and 8 parallel walkers, was utilized to get the FITS file for Model 1b. The parameters of interest are plotted with a parameter credibility interval of $\sim$95$\%$.\\
	The posterior distributions provide values consistent with the best-fit parameters listed in Table \ref{tab:pers_burst}. During the spectral fitting, the high energy cut-off ($E_{\rm cut}$) corresponding to Model 1a, and the electron temperature ($kT_{\rm e}$) corresponding to Model 1b were consistently pegged at their maximum allowed values, indicating that these parameters could not be constrained by the data. Hence, they were excluded from the corner plot.
	
	\section{Pearson correlation coefficient}
	\label{app:pears}
	\cite{pearson1895vii} measured the strength of a linear relationship between two variables $x_i$ and $y_i$ as correlation coefficient ($r$) which is given by 
	\begin{equation*}
		r = \frac{\sum_{i=1}^{n}(x_i-\bar{x})(y_i-\bar{y})}{\sqrt{\sum_{i=1}^{n}(x_i-\bar{x})^2}\sqrt{\sum_{i=1}^{n}(y_i-\bar{y})^2}}
	\end{equation*}
	where, $\bar{x}$ and $\bar{y}$ refers to the mean values of $x_i$ and $y_i$, respectively. The $r$-value is interpreted as positive, null and negative correlation for $r$=+1, 0, and -1, respectively. The corresponding $p$-value is obtained from Student's t-statistic \citep{student1908probable, peters1987pearson}. The Student's $t$-statistic is given by $t=r\sqrt{\frac{n-2}{1-r^2}}$. And, the two-sided $p$-value (significance) is  then $p=2[1-F_t(|t|,\nu)]$, where $F_t$ is the cumulative distribution function of the Student's $t$-distribution and $\nu=n-2$ is the degrees of freedom. The significance of $p$-value is that it shows statistical significance of the correlation between the variables. For $p<0.05$, the observed correlation is significant, and for $p>0.05$, the correlation is not statistically significant.\\

	\def\apj{ApJ}
	\def\apjl{ApJl}
	\def\pasp{PASP} \def\mnras{MNRAS} \def\aap{A\&A} \def\physerp{PhR} \def\apjs{ApJS} \def\pasa{PASA}
	\def\pasj{PASJ} \def\nat{Nature} \def\memsai{MmSAI} \def\araa{ARAA} \def\iaucirc{IAUC} \def\aj{AJ} \def\aaps{A\&AS} \def\ssr{SSR}
	\def\iaucirc{iaucirc}

	\bibliographystyle{elsarticle-harv}
	\let\oldbibitem\bibitem
	\renewcommand{\bibitem}{\par\oldbibitem}
	\bibliography{IGR_17062_6143_ar}
	
\end{document}